\documentclass[aps,preprint,superscriptaddress,longbibliography,nofootinbib]{revtex4-1}

\usepackage[letterpaper, margin=1in]{geometry}
\usepackage{amsmath,amsfonts,amssymb}
\usepackage{braket}
\usepackage{graphicx}
\usepackage{tikz}
\usetikzlibrary{shapes,arrows,positioning,calc}
\usepackage[hidelinks]{hyperref}

\newcommand{\mb}[1]{{\mathbf #1}}

\DeclareMathOperator{\tr}{tr}

\newcommand{\be}{\begin{equation}}
\newcommand{\ee}{\end{equation}}

\AtBeginDocument{%
    \newwrite\bibnotes
    \def\bibnotesext{Notes.bib}
    \immediate\openout\bibnotes=\jobname\bibnotesext
    \immediate\write\bibnotes{@CONTROL{REVTEX41Control}}
    \immediate\write\bibnotes{@CONTROL{%
    apsrev41Control,author="08",editor="1",pages="1",title="0",year="1"}}
     \if@filesw
     \immediate\write\@auxout{\string\citation{apsrev41Control}}%
    \fi
}%

\begin{document}

\title{Strongly incoherent gravity}

\author{Daniel Carney}
\affiliation{Physics Division, Lawrence Berkeley National Laboratory, Berkeley, CA}
\author{Jacob M. Taylor}
\affiliation{Joint Quantum Institute, NIST/University of Maryland, College Park, MD}

\date{\today}

\begin{abstract}
While most fundamental interactions in nature are known to be mediated by quantized fields, the possibility has been raised that gravity may behave differently. Making this concept precise enough to test requires consistent models. Here we construct an explicit example of a theory where a non-entangling version of an arbitrary two-body potential $V(r)$ arises from local measurements and feedback forces. While a variety of such theories exist, our construction causes particularly strong decoherence compared to more subtle approaches. Regardless, expectation values of observables obey the usual classical dynamics, while the interaction generates no entanglement. Applied to the Newtonian potential, this produces a non-relativistic model of gravity with fundamental loss of unitarity. The model contains a pair of free parameters, a substantial range of which is not excluded by observations to date. As an alternative to testing entanglement properties, we show that the entire remaining parameter space can be tested by looking for loss of quantum coherence in small systems like atom interferometers coupled to oscillating source masses.

\end{abstract}

\maketitle

\section{Introduction}

The past ten years have seen an explosion of proposals \cite{Carney:2018ofe,kafri2013noise,Kafri:2014zsa,kafri2015bounds,Bahrami:2015wma,anastopoulos2015probing,bose2017spin,marletto2017gravitationally,haine2021searching,Qvarfort:2018uag,Carlesso:2019cuh,PhysRevA.101.063804,howl2021non,Matsumura:2020law,Pedernales:2021dja,liu2021gravitational,Datta:2021ywm}, building on earlier conceptual suggestions \cite{cecile2011role,feynman1971lectures,page1981indirect}, to test whether or not gravitational interactions can entangle objects. When general relativity is quantized in perturbation theory, in direct analogy with quantum electrodynamics or any of the other known field theories in nature \cite{Donoghue:1994dn,Donoghue:1995cz,Burgess:2003jk}, non-relativistic gravitational interactions do generate entanglement \cite{Carney:2018ofe,Carney:2021vvt}. It is thus of crucial importance to develop theoretically consistent models of the alternative hypothesis, in which gravity \emph{cannot} entangle objects, in order to understand what these entanglement experiments may be able to rule out.

Models where non-entangling (``classical'') gravity interacts with quantum matter take a variety of forms; a non-exhaustive set of examples includes  \cite{kafri2013noise,Kafri:2014zsa,kafri2015bounds,tilloy2016sourcing,Carney:2018ofe,Hall:2017nzl,Tilloy:2018tjp,Oppenheim:2018igd,Grossardt:2022zsi,Giulini:2022wyp,Layton:2022sku,ma2022limits}. Typically, these involve a classical gravitational field $h_{\mu\nu}$ coupled to matter like $H_{\rm int} \sim h_{\mu\nu} \braket{T^{\mu\nu}_{\rm matter}}$. To avoid fundamental causality problems \cite{polchinski1991weinberg}, this usually requires that the effective quantum dynamics of the matter is non-unitary \cite{Carney:2018ofe}. This leads to a variety of observable signatures: entanglement is not generated, energy-momentum conservation is violated \cite{banks1984difficulties}, and quantum coherence can be lost. While entanglement generation offers a clean interpretation, the other effects may be more amenable to near-term experiments, and therefore merit detailed study.

In this paper, we focus on the loss of quantum coherence of matter these scenarios. To make the issue precise, we develop a model in which the non-relativistic gravitational interaction arises as an error process, building on prior work on a lattice~\cite{Kafri:2014zsa,kafri2015bounds}. Random position measurements act on the matter, much as in local spontaneous collapse models \cite{bassi2013models}, and the measurement results are used to generate a noisy gravitational force. Classical gravitational interactions arise in the limit of large unentangled masses, but new effects appear in the quantum regime. In contrast to general measure-and-feedback models of gravity (e.g.,~\cite{kafri2013noise,Oppenheim:2018igd}), this theory has measurements that are strong and feedback that is weak, and is thus ``strongly incoherent''. This incoherence means that observing the persistence of specific coherences enables one to rule out the entire parameter space in a direct fashion, such as by using the quantum collapse-and-revival experiment proposed in \cite{Carney:2021yfw} (see also \cite{Streltsov:2021ahn}). In particular, this experiment is substantially easier than a direct test of entanglement generation, and could be performed by studying interactions between a state-of-the-art atom interferometer and high-quality mechanical system \cite{hamilton2015atom,xu2019probing,Panda:2022gtw}.

\section{Construction of the model}
\label{sec-general}

Our aim is to find a time evolution law on a non-relativistic $N$-body quantum system which realizes a ``classical'' version of some two-body potential $V(\mb{x}_i - \mb{x}_j)$ on the particles. By classical version of the potential, we mean something where no entanglement can be generated, but where the average particle motion obeys the usual semiclassical version of the Ehrenfest limit\footnote{We will have many equations that use both the position operator $\hat{\mb{x}}$ and its eigenvalues $\mb{x}$, so we will put hats on the position operators for clarity. We also note that bold-face variables are 3-vectors. Other operators, for example $\mb{p}$, will remain hatless. We use $\hbar = c = k_B = 1$ units, returning to regular SI units when computing explicit bounds.}
\be
\label{ehrenfest}
\frac{d\braket{\mb{p}_i}}{dt} = -i \braket{ [H, \mb{p}_i]} -\sum_{j \neq i} \Braket{ \nabla V (\hat{\mb{x}}_i - \hat{\mb{x}}_j)}\ .
\ee
In the limit that the total state is approximately non-entangled and particles are distant, $\braket{V(\hat{\mb{x}}_i - \hat{\mb{x}}_j)} \approx V(\Braket{\hat{\mb{x}}_i} - \Braket{\hat{\mb{x}}_j})$ can be realized, and we see a semiclassical limit emerge. Here $H$ means the Hamiltonian other than the potential interaction $V$ of interest. In the gravitational context $V = -G_N m_i m_j/|\hat{\mb{x}}_i - \hat{\mb{x}}_j|$, this limit means that the model can accurately reproduce previously observed classical gravitational phenomena like the orbit of the Earth around the Sun.

As described above, the model consists of random ``errors'' which involve both a measurement and an injection of energy. First, we describe the measurement process. Consider a measurement of a particle's position $\hat{\mb{x}}$ with some uncertainty $\sigma$. Since this is a non-projective measurement, it is described by a positive operator-valued measurement (POVM)
\be
\label{completeness}
\int d^3\mb{x} P^\dag(\mb{x}) P(\mb{x}) = 1,
\ee
where for example
\be
\label{gaussianpovm}
P(\mb{x}) = (2\pi \sigma^2)^{-3/4} \exp \left\{ -(\mb{x}-\hat{\mb{x}})^2/4 \sigma^2 \right\}
\ee
represents the outcome that the particle's position is $\hat{\mb{x}} = \mb{x}$ within a Gaussian error of $\sigma$. The probability of outcome $\mb{x}$ is, as usual, $p(\mb{x}) = \braket{ P^\dag(\mb{x}) P(\mb{x})}$. The precise form of the POVM is not important in what follows.

The classical record of the outcome $\mb{x}$ can then be fed back onto the other particles in the form of a force. Given outcome $\mb{x}_i$ for the $i$th particle, we can act on another particle $j \neq i$ with coordinate $\hat{\mb{x}}_j$ through the operator
\be
U_{ij}(\mb{x}_i - \hat{\mb{x}}_j) = \exp \left\{ -i \eta_{ij} \phi(\mb{x}_i - \hat{\mb{x}}_j) \right\}.
\ee
Notice that $U$ is a unitary operator that acts \emph{only} on particle $j$, using the classical record of the other particle's position $\mb{x}_i$. The ``coupling constants'' $\eta_{ij}$ and so-far arbitrary potential function $\phi(\mb{x}-\mb{y})$ will be used to mimic the desired classical potential; we will discuss them in detail shortly.

With these two elements, we can now describe the total time evolution. Define the Lindblad (jump) operators
\be
\label{errors}
E_i(\mb{x}) = P_i(\mb{x}) \prod_{j \neq i} U_{ij}(\mb{x} - \hat{\mb{x}}_j).
\ee
This operator measures particle $i$ and applies a kick to each other particle with potential determined by the outcome $\mb{x}_i = \mb{x}$. The total time evolution is then taken to be of Lindblad form, with the $E_i$ used as the Lindblad operators:
\begin{align}
\label{lindblad}
\dot{\rho} = -i [H,\rho] + \sum_i \gamma_i \int d^3\mb{x} \Big[ E_i(\mb{x}) \rho E^\dag_i(\mb{x})  - \frac{1}{2} \left\{ E^\dag_i(\mb{x}) E_i(\mb{x}), \rho \right\} \Big].
\end{align}
Here we have neglected the Hamiltonian terms for simplicity. The parameters $\gamma_i$ are rates which describe how often the measurement-feedback operation occurs. By construction, this structure is explicitly LOCC, and thus implies that the interactions cannot generate any entanglement. In addition, this is local in time, preserves the density matrix norm $\tr \rho = 1$, each Lindblad operator is separable $E_i = \mathcal{O}_1 \otimes \mathcal{O}_2 \otimes \cdots$, and the effective Hamiltonian terms $E_i^\dag E_i$ operate only on particle $i$. This latter set of constraints is why we call this `strongly incoherent', and enables our revival test in the latter half of the paper. Note however that this does not rule out other proposed non-entangling theories, just the category of 'strongly incoherent ones' introduced here.

\begin{figure}[t]
\includegraphics[scale=1.2]{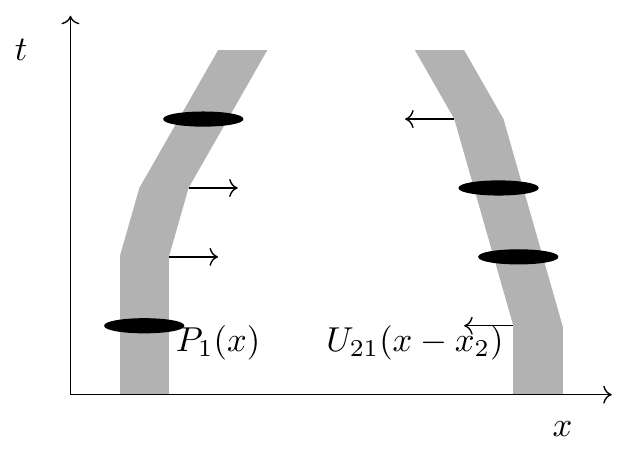}
\caption{Schematic time evolution of a pair of massive particles under strongly incoherent gravitational interactions. The average trajectories obey the usual classical equations of motion, but there are random deviations caused by both the position measurements $P(x)$ and noisy gravitational kicks $U(x)$. The grey bands represent position-space wavefunctions and the black discs represent insertions of the measurement operators.}
\label{fig-cartoon}
\end{figure}

The time evolution \eqref{lindblad} represents a series of random measurements and kicks, leading to an overall motion of the particles; see Fig. \ref{fig-cartoon}. Notice that the entire process is local in the sense of Hilbert space, but non-local in the sense of spacetime. Both the measurement and feedback happen instantaneously and at spacelike separation. Since our goal is to reproduce a non-relativistic potential interaction, this is acceptable. The model in this non-relativistic form has interactions which are explicitly dependent on single-particle locations through the $P_i$ operators, and thus breaks translation invariance. Further, the jump operators $E_i$ do not commute with the particle kinetic energy terms, so time translation invariance is also broken. Thus the model will have anomalous heating and dephasing effects, as discussed in the next section.

Finally, we want to impose our requirement that the averaged dynamics obey semiclassical equations of motion like \eqref{ehrenfest} in the limit that the initial states are nearly classical. For an un-entangled $N$-body state $\rho = \rho_1 \otimes \rho_2 \otimes \cdots$, the correlation function factors, and we can insert a complete set of states in \eqref{ehrenfest} to write a given interaction term as
\be
\label{ehrenfest2}
\Braket{\nabla V (\hat{\mb{x}}_i - \hat{\mb{x}}_j)} = \int d^3\mb{x} \Braket{\nabla V (\hat{\mb{x}}_i - \mb{x})} \rho_j(\mb{x}).
\ee
Here $\rho_j(\mb{x}) = \braket{\mb{x} | \rho_j | \mb{x}}$ are the diagonal elements of the $j$th particle's position-space density matrix. We want to reproduce this in the same limit in our strongly incoherent interaction model. Using our time evolution \eqref{lindblad}, and ignoring the Hamiltonian terms for simplicity, direct computation using the Heisenberg-picture Lindblad equation gives
\begin{align}
\begin{split}
\label{dpdt}
\frac{d\braket{\hat{\mb{p}}_i}}{dt} & = -\sum_{j \neq i} \eta_{ij} \gamma_j \int d^3\mb{x} \Braket{\nabla \phi (\mb{x} - \hat{\mb{x}}_i) P_j^\dag(\mb{x}) P_j(\mb{x}) } \\
& \approx -\sum_{j \neq i} \eta_{ij} \gamma_j \int d^3\mb{x} \Braket{\nabla \phi (\mb{x} - \hat{\mb{x}}_i)} p_j(\mb{x})
\end{split}
\end{align}
where the first equality is exact, and the second line follows in the limit that the $N$-body state is unentangled. Here, we identified the quantity
\be
p_j(\mb{x}) = \Braket{ P_j^\dag(\mb{x}) P_j(\mb{x})} \label{e:prob}
\ee
as a probability distribution for the position of particle $j$. This is just a noisy representation of the diagonal elements of the density matrix $p_j(\mb{x}) \approx \rho_j(\mb{x})$ for particle $j$, with errors of order $\sigma$. Comparing to \eqref{ehrenfest2}, we see that to reproduce a given two-body potential of the form $V(\mb{x}_i - \mb{x}_j) = \alpha_{ij} \phi(\mb{x}_i - \mb{x}_j)$, we need to identify the coupling constants
\be
\alpha_{ij} = \eta_{ij} \gamma_j.
\ee
With these identifications, and assuming that our position measurement errors $\sigma$ are small enough to give a good approximation to the particle density matrix $p_j(\mb{x}) \approx \rho_j(\mb{x})$, our noisy equation of motion \eqref{dpdt} reduces to the correct Ehrenfest limit \eqref{ehrenfest}.

Let us now apply this construction to Newtonian gravity. In this case, the couplings $\alpha_{ij} = -G_N m_i m_j$ and $\phi(\mb{x}_i-\mb{x}_j) = 1/|\mb{x}_i - \mb{x}_j|$. Thus we need to choose the coupling parameters $\eta_{ij}, \gamma_i$ so that their product gives $\eta_{ij} \gamma_j = G_N m_i m_j$ (no sum on the $j$ index). The factor $\gamma_i$ has units of a rate, so we can take it to be proportional to the appropriate mass. This fixes the $\eta_{ij}$ to be $i$-independent, and we can write
\be
\label{vsigma}
\gamma_i = v^2 m_i, \ \ \ \eta_{ij} = G_N m_j/v^2,
\ee
where $v$ is a dimensionless real parameter that can be thought of as a velocity in units of $c$, the speed of light. This ability to rescale the two couplings by reciprocal powers of $v$ reflects a physical symmetry of the model. Our demand is only that the average equations of motion look like Newtonian gravity. This can be accomplished by either many rapidly applied weak kicks (large $v$), or rare but strong kicks (small $v$). Thus in total, our ``strongly incoherent gravity'' model depends on two free parameters: $v$ and the position measurement error $\sigma$. In Sec. \ref{sec-gravity}, we will show how to use a variety of existing experimental data to constrain a large range of these parameters; in Sec. \ref{sec-tests}, we show how to test the currently unconstrained parameter space.

\begin{figure}[t]
\includegraphics{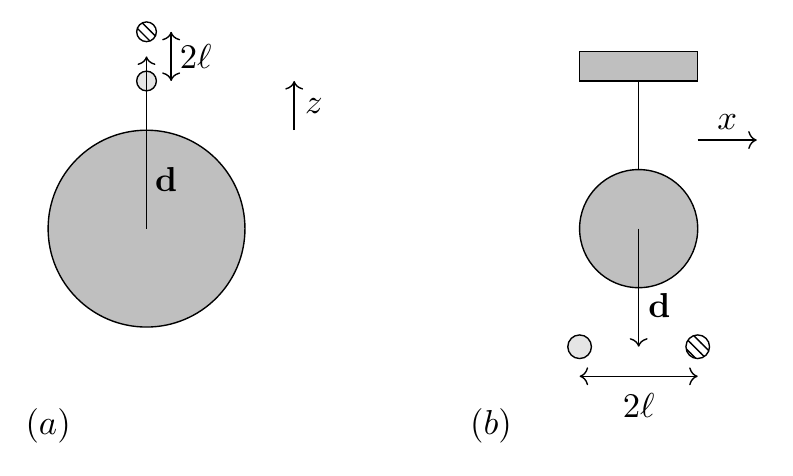}
\caption{Two geometries of interest, with a small mass $m$ in superposition of two localized states $\ket{L}, \ket{R}$ near a large mass $M$. (a) The dipole interaction leads to a classical relative phase shift from a stationary heavy mass (Sec. \ref{sec-general}). (b) Here $M$ is suspended as a movable pendulum, and the quadrupole interaction between $M$ and $m$ can lead to either entanglement or noisy gravitational interactions, depending on the gravity model under consideration (Sec. \ref{sec-tests}).}
\label{fig-geometries}
\end{figure}

Before moving to the tests, it is instructive to see in detail how our strongly incoherent gravity model reproduces the predictions of classical gravity, as well as the standard quantum Newton potential, in a situation involving non-trivial quantum systems. Consider placing a large source mass $M$ in a fixed location, and positioning a small mass $m$ nearby, with $m$ prepared in an initial superposition of two well-localized states $(\ket{L} + \ket{R})/\sqrt{2}$ separated by a distance $\ell$. This small mass could be, for example, a neutron in an interferometer placed above the Earth \cite{colella1975observation}, or a fountain atom interferometer next to a kg-scale mass \cite{overstreet2022observation}. We can approximate the position operator of the small mass as
\be
\hat{\mb{x}} = \mb{d} + \mb{\ell} \mb{e}_z \sigma_z/2,
\ee
where $\sigma_z = \ket{L}\bra{L} - \ket{R} \bra{R}$, $\mb{d}$ is the vector from the center of the two sites to the source mass, and $\mb{e}_z$ is a unit vector (see Fig. \ref{fig-geometries}). We will be interested in the behavior of the fringe coherence $\braket{\sigma_-(t)}$, where $\sigma_- = \ket{L} \bra{R}$ measures the off-diagonal component of the density matrix.

To understand the basic idea, we can first consider a heuristic model of a classical source mass $M$ coupled to the atom, following \cite{Hosten:2021biw}. Consider taking the source mass position as a fixed, classical value, so that $\mb{d}$ is just a number. This generates a simple external Newton potential on the atom. We can expand the Newton potential in multipoles [see Eq. \eqref{multipole}]. The monopole term acts identically on both branches and gives an overall phase. The first non-trivial effect comes from the dipole term, which generates the evolution
\be
\ket{L} + \ket{R} \to e^{-i \varphi(t)} \ket{L} + e^{+i \varphi(t)} \ket{R}, \ \ \varphi(t) = -\frac{G_N M m \ell t}{2 d^2},
\ee
so the coherence evolves as
\be
\label{classical-main}
\braket{\sigma_-(t)} = e^{-2 i \varphi(t)} \braket{\sigma_-(0)}.
\ee
This explains the observed oscillating interference fringes with contrast proportional to $G_N$ in these experiments \cite{colella1975observation,overstreet2022observation}. When the Newton potential is instead treated as an entangling two-body operator, and the source $M$ is taken to be very heavy $M \gg m$ in a well-localized state, precisely the same coherent evolution occurs. Thus these types of experiments cannot distinguish between this heuristic classical model and the usual quantum Newtonian potential that arises in effective field theory.

We can now contrast this to our strongly incoherent gravity model. Crucially, although this model is designed to reproduce semi-classical gravity, it requires treating the source mass $M$ as a quantum system. As above, the first non-trivial effect comes from the dipole term in the potential function $\phi$. The ``kick'' operator $U$ acting on the light mass in the error operators \eqref{errors}, to this order, is $U = \exp(-i G_N m \ell \sigma_z/2 v^2 d^2)$. Using this in the Heisenberg picture version of \eqref{lindblad}, the equation of motion for $\braket{\sigma_-(t)}$ is
\be
\label{sigmadot-main}
\braket{\dot{\sigma}_-} = \gamma_M \left[ e^{i G_N m \ell/v^2 d^2} - 1 \right] \braket{\sigma_-} = i \frac{G_N M m \ell}{d^2} \braket{\sigma_-} + \mathcal{O}(G_N^2).
\ee
Here we used $\gamma_M = v^2 M$, the completeness of the POVM \eqref{completeness}, and ignored the dephasing term $\sim \gamma_m = v^2 m$ from the error operators that measure the small mass position. The powers of our free parameter $v^2$ have cancelled, and this result shows that the coherence oscillates precisely as in the classical and standard quantum models above, c.f. equation \eqref{classical-main}. To see how the strongly incoherent gravity model differs, we will need to go to the next order in the interaction, where noise effects become visible.

\section{Constraints on the free parameters}
\label{sec-gravity}

We now turn to understanding constraints on the possible values of the free parameters of our model, $\sigma$ and $v$. By construction, the model reproduces classical gravity at the level of average values for observables. Therefore, we want to study deviations from the averages in order to look for constraints. In particular, as discussed above, the noisy gravitational interaction will produce anomalous heating and dephasing. These effects are strongly constrained by various precision tests of cold and quantum coherent systems. We show the totality of these bounds in Fig. \ref{fig-bounds}.

\subsection{Spontaneous collapse effects}

In our strongly incoherent gravity model, each massive object is occasionally measured with the operators $P(\mb{x})$, regardless of whether the mass is interacting with another object. This means that our model has ``spontaneous collapse'' or localization effects. In particular, the measurement process will both collapse spatial superpositions and generate random momentum kicks, even for a completely isolated particle.

We begin with the collapse of superpositions. Consider a pair of position eigenstates $\ket{\mb{x}_L}, \ket{\mb{x}_R}$ separated by a distance $\Delta x = |\mb{x}_L - \mb{x}_R|$. The density matrix element $\rho_{LR} = \braket{\mb{x}_L | \rho | \mb{x}_R}$ quantifies the coherence of the superposition of these two states. Taking the appropriate matrix element in \eqref{lindblad}, we see that this coherence evolves as
\be
\frac{d\rho_{LR}}{dt} = \gamma \left[ e^{-\Delta x^2/8 \sigma^2} - 1 \right] \rho_{LR} \approx - \gamma \rho_{LR}
\ee
where the approximation is good for superpositions larger than the measurement length $\Delta x \gtrsim \sigma$. Thus a spatial superposition of a particle is decohered at the rate $\gamma = v^2 m$. Comparing to the observed $\sim 60~{\rm s}$ coherence times in atom interferometers \cite{xu2019probing,Panda:2022gtw} then places an upper bound on $v^2$, for all $\sigma \lesssim 1~{\rm cm}$, the separation scale in these experiments.

The momentum kicks lead to anomalous heating effects. Consider the action of the position measurement $P(\mb{x})$ on a momentum eigenstate $\ket{\mb{p}}$. With the explicit Gaussian POVM \eqref{gaussianpovm}, the resulting post-measurement state $P(\mb{x}) \ket{\mb{p}}$ is a Gaussian of width $\Delta p_{\rm kick} = 1/2 \sigma$, assuming $\sigma$ is smaller than the initial wavefunction scale. Thus the action of $P(\mb{x})$ on a generic initial state will increase the momentum uncertainty of the particle by at least $\Delta p_{\rm kick}$, or in other words it will generate an RMS energy of order
\be
\Delta E_{\rm kick} \approx \frac{1}{m \sigma^2}.
\ee
These kicks happen at a rate $\gamma = v^2 m$.

In the limit that the kicks are strong but rare (small $v$), strong bounds can be derived by consider some prototypical atomic systems. If the typical kick exceeds the depth of a trapped atom $\Delta E_{\rm kick} \gtrsim E_{\rm trap}$, then the will be ejected roughly once per $1/\gamma$. In a free-falling atom interferometer, a kick even of order $\Delta E_{\rm kick} \gtrsim E_{\rm int} \approx k$, where $k$ is the Raman momentum, will displace the atom from the interference region and effectively look like a loss. In either case, we can place a very conservative bound by demanding that in an experiment of duration $\tau \gtrsim 1/\gamma = 1/v^2 m_a$, with $m_a$ the atomic mass, on average no atoms will be lost, $\Delta E_{\rm kick} \lesssim E_{\rm trap}, E_{\rm int}$. In Fig. \ref{fig-bounds} we plot these bounds based on state of the art atomic experiments. 

In the opposite regime of large $v$, where the kicks are weak but happening very often, one needs to instead consider continuous anomalous heating effects. This can be quantified by using the Heisenberg picture version of \eqref{lindblad} to compute, for a single isolated particle, the RMS change in momentum:
\begin{align}
\label{dpdtba}
\Braket{ \frac{d\Delta \mb{p}^2}{dt}}_{\rm BA} & = v^2 m \int d^3\mb{x} \Braket{ \left( \nabla P(\mb{x}) \right)^2} \approx \frac{v^2 m}{4 \sigma^2},
\end{align}
where the final estimate comes from using the explicit Gaussian measurement \eqref{gaussianpovm}. The subscript ``BA'' indicates that this is backaction noise from the measurement process. This can be viewed as an anomalous heating
\be
\label{dedt}
\Braket{ \frac{dE_{\rm rms}}{dt}}_{\rm BA} \approx \frac{v^2}{\sigma^2} \approx 0.6~\frac{\rm \mu K}{\rm s} \times \left( \frac{v}{10^{-15}} \right)^2 \left( \frac{1~{\rm nm}}{\sigma} \right)^2,
\ee
independent of the particle mass. We can compare this to a large mass composed of $N$ atoms placed in a dilution refrigerator \cite{bassi2013models}. These systems have typical cooling power $P_{\rm cool} \approx 10~{\rm \mu W}$ \cite{zu2022development}, which corresponds to $\braket{dE_{\rm rms}/dt}_{\rm cool} \approx P_{\rm cool}/N$. This corresponds to an upper bound on $v/\sigma$, the diagonal region in Fig. \ref{fig-bounds}, where we used a kilogram-scale copper mass as a benchmark. One can also consider bounds from smaller systems like trapped ions or Bose-Einsten condensates, which give similar bounds.

\subsection{Noise in the gravitational interaction}

\begin{figure}[t!]
\includegraphics[scale=0.53]{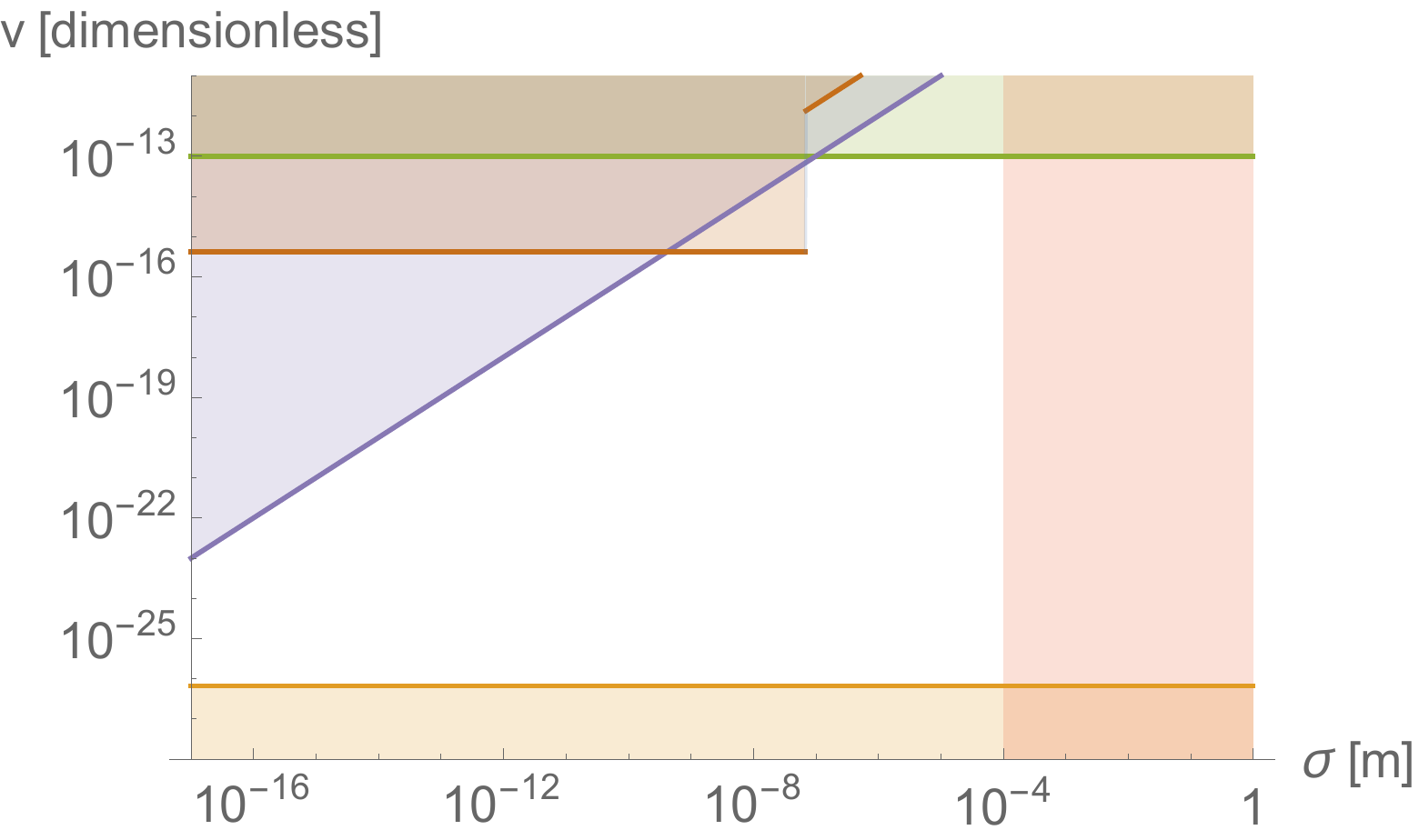}
\caption{Constraints on the parameters $v$ and $\sigma$. Lack of single lost atoms leads to a bound on the backaction kick in cold atom traps (red, top left, \cite{scherschligt2017development}) and long-hold Penning traps (green, top, \cite{gabrielse1999precision}). Bounds on position-dephasing driven heating arise from observed cooling capabilities of dilution refrigerators (purple, \cite{zu2022development}). Constraints on the shot noise in the gravitational interaction arise from a lack of anomalous heating in interferometry experiments near the Earth or other large mass (orange, bottom region, \cite{xu2019probing,overstreet2022observation,Panda:2022gtw}). Finally, short-range observations of Newton's gravitational force law preclude $\sigma > 100~{\mu m}$ (red, right region, \cite{wagner2012torsion}). The resulting white region appears to be allowed by current experimental constraints.
}
\label{fig-bounds}
\end{figure}

In the previous section, we quantified the effects arising entirely from the spontaneous position measurements in the model. Now we will consider the emergent gravitational interaction. This is no longer coherent but has a degree of shot noise, because the gravitational force (the $U_{ij}$ operators) is sourced by a noisy estimate of the source position. 

Again using the Heisenberg-picture Lindblad equation, one finds by a direct but somewhat lengthy computation that the variance $\Delta \mb{p}_i^2$ of the $i$th particle's momentum will increase over time, according to
\be
\frac{d \Delta \mb{p}_i^2}{dt} =  \left(\frac{d \Delta \mb{p}_i^2}{dt}\right)_{\rm BA} + \left(\frac{d \Delta \mb{p}_i^2}{dt}\right)_{\rm shot}.
\ee
The first term is given by \eqref{dpdtba}, and the second term comes from the noisy interaction with the other particles:
\begin{align}
\label{heating}
\Braket{\frac{d \Delta \mb{p}_i^2}{dt}}_{\rm shot} = \frac{G_N^2 m_i^2}{2 v^2} \int d^3\mb{x} \rho_{m,i}(\mb{x}) \braket{ \left( \nabla \phi(\mb{x}-\mb{x}_i) \right)^2}.
\end{align}
Here, $\rho_{m,i}(\mb{x}) = \sum_{j \neq i} m_j p_j(\mb{x})$ is a noisy estimate of the mass density of all the particles $j \neq i$, and $\phi(\mb{x}) = 1/|\mb{x}|$ as before. Since we will apply this computation only in the semiclassical limit, we have assumed that the $N$-body state is unentangled as before, and also that the effective potential on the $i$th particle is approximately constant across its spatial wavefunction. See Appendix \ref{appendix-heating} for a detailed calculation.

Consider the effects of this gravitational shot noise on a small system near a large source mass, like an atom interferometer some distance $d$ above the surface of the Earth. As the $(\nabla \phi)^2$ term goes as $1/r^4$, it is this nearest distance that matters when $d$ is much less than the radius of the Earth. One can estimate the integral in \eqref{heating} to obtain a heating rate for the atom of order
\be
\label{dedt-numbers}
\Braket{ \frac{d \Delta \mb{p}^2}{dt}}_{\rm shot} \approx \frac{G_N^2 m_a^2 \rho_0}{v^2 d}
\ee
where $\rho_0 \approx 1800~{\rm kg/m^3}$ is the solid density of the Earth and $m_a$ is the atom's mass. A constant heating rate will lead to position uncertainty in the atom of order $\Delta x^2 \approx (\tau^3/m^2) (d \Delta p^2/dt)$ after a time $\tau$. This cannot be larger than the de Broglie wavelength $\lambda_{dB}$ of atoms in the interferometer without destroying phase contrast, leading to a \emph{lower} bound on the parameter $v$:
\be
\Delta x^2 \approx \frac{G_N^2 \rho_0 \tau^3}{v^2 d} \lesssim \lambda_{dB}^2.
\ee
In Fig. \ref{fig-bounds}, we display this bound assuming an atom interferometer using rubidium $m_a = 85~{\rm u}$ placed $d = 1~{\rm m}$ above the Earth and held for $\tau \approx 20~{\rm s}$ \cite{xu2019probing,overstreet2022observation,Panda:2022gtw}.

\subsection{Modified Newtonian gravity}

In addition to the noise in the Newtonian interaction, the average DC potential itself will be modified at short distances $\Delta x \lesssim \sigma$ in this strongly incoherent gravity model. The nominal particle locations are only accurate to around $\sigma$, so the effective potential will like convolution of the Newtonian force law with a distribution function of the particles' locations. Specifically, consider the effective potential appearing in \eqref{ehrenfest2},
\be
\braket{V_{{\rm eff}}(\hat{\mb{x}}_i)} \approx -\sum_{i} G_N m_i m_j \int d^3\mb{x} \Braket{\frac{1}{\left| \hat{\mb{x}}_i - \mb{x} \right|}} p_j(\mb{x}).
\ee
The probability distribution $p_j(\mb{x})$ varies non-trivially on the measurement distance scale $\sigma$. Thus, if mass $i$ is measured at a distance around $\sigma$ or less from the other particles $j \neq i$, the effective potential will be significantly modified, and in particular the short-distance singularity in the potential would be ``softened''. Similar effects in astrophysics are often modelled by potentials of the form $V_{\rm eff} \sim 1/{\sqrt{|r|^2+\sigma^2}}$ \cite{barnes2012gravitational}. Searches for modifications to the Newton potential currently operate around $100~{\mu m}$ and above and are consistent with an unmodified potential \cite{wagner2012torsion}. Demanding that these effects are not present then sets a conservative upper bound on the allowed values of $\sigma$, as displayed in Fig. \ref{fig-bounds}.

\section{Direct experimental tests}
\label{sec-tests}

As described in the previous section, there is a large range of model parameters $v,\sigma$ for which the strongly incoherent gravitational interaction appears to be viable. Testing the remainder of the parameter space could be done by further refining the kinds of experiments discussed there, for example, more precise anomalous heating measurements. A more direct set of tests arise in situations where gravitationally-coupled objects are prepared in non-trivial quantum states. A straightforward approach is to observe entanglement generated between two massive objects via their gravitational interaction \cite{kafri2013noise,Kafri:2014zsa,Bahrami:2015wma,anastopoulos2015probing,bose2017spin,marletto2017gravitationally,haine2021searching,Qvarfort:2018uag,Carlesso:2019cuh,PhysRevA.101.063804,howl2021non,Matsumura:2020law,Pedernales:2021dja,liu2021gravitational,Datta:2021ywm}. Observation of this entanglement would entirely rule out our strongly incoherent gravity model, which cannot entangle objects. Current proposals for these experiments, however, are very challenging, and may take decades to complete. A more straightforward alternative would be to look for the \emph{lack} of the loss of quantum coherence predicted in a small system coupled to a gravitational source due to the inherent noise in the incoherent interaction.

Consider again a heavy source mass $M$ near a spatially superposed light mass $m$, as discussed at the end of Sec. \ref{sec-general}. There, we showed that the lowest-order gravitational effect, namely the dipole-induced coherent phase shift, arises identically in the heuristic classical model, in standard quantized gravity with a coherent entangling Newton potential, and in our strongly incoherent gravity model. 

Let us now consider the leading corrections. In the heuristic classical model of \cite{Hosten:2021biw}, the prediction \eqref{classical-main} is exact. In both standard quantized gravity and in our strongly incoherent gravity model, the source $M$ is a quantum object. In the former case, with a coherent entangling Newton potential, \eqref{classical-main} is accurate up to extremely small corrections coming from standard decoherence. In our strongly incoherent gravity model, however, there is a substantial correction which leads to loss of coherence in the superposition. We need to specify a quantum state for the heavy mass; we can approximate the static classical situation by assuming $M$ is in a highly localized state, for example, $\psi_M(z_M) = N e^{-z_M^2/4 \Delta z_M^2}$ with small uncertainty $\Delta z_M \to 0$. We will focus only on the $z$ axis in what follows. The first non-trivial corrections to \eqref{sigmadot-main} arise from the quadrupole term in the potential of the form $\phi_{qp} = 2 z_m z_M/d^3$ [c.f. equation \eqref{multipole}]. This term leads to an $\mathcal{O}(G_N^2)$ correction, namely
\be
\label{loss}
\braket{\dot{\sigma}_-} = \left[ i \frac{G_N M m \ell}{d^2} - \frac{1}{v^2 M} \left( \frac{G_N M m \ell \sigma}{d^3} \right)^2 + \mathcal{O}(G_N^3) \right] \braket{\sigma_-},
\ee
which can be solved easily,
\be
\label{sigmadotincoh}
\braket{\sigma_-(t)}_{\rm incoh} = e^{-2 i \varphi(t)} e^{-\Gamma_{\rm incoh} t}, \ \ \ \Gamma_{\rm incoh} =  \frac{G_N^2 M m^2 \ell^2 \sigma^2}{d^6 v^2}.
\ee
Here the first term leads to the same coherent oscillation as in \eqref{classical-main}, $\sigma$ is the measurement length scale of the model, and we assumed $\sigma \gg \Delta z_M$. Notice in particular that this decoherence rate $\sim 1/v^2 \gg 1$. We provide a detailed calculation of \eqref{loss} in Appendix \ref{appendix-visibility}, but the essential physics is clear from the presence of $\sigma$ in the exponent: this decoherence comes from the shot noise in the gravitational interaction. Observation of atomic coherence longer than $1/\Gamma_{\rm incoh}$ would then rule out this model.

Finally, we remark on a much cleaner test, which can definitively probe the entire unconstrained parameter space of strongly incoherent gravity. If the source mass $M$ has periodic behavior, like a suspended pendulum as depicted in Fig. \ref{fig-geometries}, then one can look for the periodic quantum collapse and revival of the atomic state due to the periodic dynamics of the pendulum. Observationally, this manifests as a visibility $V(t)$ which oscillates sinuisoidally with the frequency $\omega_M$ of the source. This signature was studied in detail in \cite{Carney:2021yfw}, where it was shown that the revival occurs assuming the usual quantum-coherent gravitational interaction, even if the oscillator is in a high temperature state. The same revival also occurs in the simple heuristic classical model discussed above, as emphasized in \cite{Hosten:2021biw} (see also \cite{ma2022limits}). However, as one can see from \eqref{sigmadotincoh}, our strongly incoherent gravity model predicts $\dot{V} < 1$. In Appendix \ref{appendix-visibility}, we show that this effect persists with an oscillator and/or with initial entanglement between $M$ and $m$. Thus, observation of any non-negativity in the visibility---i.e., any level of atomic revival---would rule out the entire parameter space of this model.

\section{Outlook}
\label{sec-outlook}

All non-entangling models of gravity can be ruled out by a direct observation of gravitational entanglement generation. The results of this paper suggest that a much more immediately available goal is to rule at least some of them out through simpler quantum coherence measurements. In particular, the strongly incoherent gravity model presented in this paper leads inevitably to a complete non-revival of atomic visibility in an experiment like that proposed in \cite{Carney:2021yfw}. A similar but weaker conclusion can be drawn in other known classical, non-entangling models of gravity, which all appear to feature anomalous heating effects, leading to partial loss of quantum coherence. We leave a general classification of non-entangling models and their incoherence signatures to future work. Our strongly incoherent gravity model should provide a useful benchmark for these experiments, as well as a helpful theoretical construction which can be built out into more sophisticated, and ideally relativistic, models of non-entangling gravity.

\section*{Acknowledgements}

We thank Scott Glancy, Alexey Gorshkov, and Igor Pikovski for comments on this paper, and Julen Pedernales, Martin Plenio, and Kirill Streltsov for correspondence regarding related ideas.  DC is supported by the US Department of Energy under contract DE-AC02-05CH11231 and Quantum Information Science Enabled Discovery (QuantISED) for High Energy Physics grant KA2401032.

\bibliography{strongly-incoherent-gravity.bib}

\begin{thebibliography}{48}%
\makeatletter
\providecommand \@ifxundefined [1]{%
 \@ifx{#1\undefined}
}%
\providecommand \@ifnum [1]{%
 \ifnum #1\expandafter \@firstoftwo
 \else \expandafter \@secondoftwo
 \fi
}%
\providecommand \@ifx [1]{%
 \ifx #1\expandafter \@firstoftwo
 \else \expandafter \@secondoftwo
 \fi
}%
\providecommand \natexlab [1]{#1}%
\providecommand \enquote  [1]{``#1''}%
\providecommand \bibnamefont  [1]{#1}%
\providecommand \bibfnamefont [1]{#1}%
\providecommand \citenamefont [1]{#1}%
\providecommand \href@noop [0]{\@secondoftwo}%
\providecommand \href [0]{\begingroup \@sanitize@url \@href}%
\providecommand \@href[1]{\@@startlink{#1}\@@href}%
\providecommand \@@href[1]{\endgroup#1\@@endlink}%
\providecommand \@sanitize@url [0]{\catcode `\\12\catcode `\$12\catcode
  `\&12\catcode `\#12\catcode `\^12\catcode `\_12\catcode `\%12\relax}%
\providecommand \@@startlink[1]{}%
\providecommand \@@endlink[0]{}%
\providecommand \url  [0]{\begingroup\@sanitize@url \@url }%
\providecommand \@url [1]{\endgroup\@href {#1}{\urlprefix }}%
\providecommand \urlprefix  [0]{URL }%
\providecommand \Eprint [0]{\href }%
\providecommand \doibase [0]{http://dx.doi.org/}%
\providecommand \selectlanguage [0]{\@gobble}%
\providecommand \bibinfo  [0]{\@secondoftwo}%
\providecommand \bibfield  [0]{\@secondoftwo}%
\providecommand \translation [1]{[#1]}%
\providecommand \BibitemOpen [0]{}%
\providecommand \bibitemStop [0]{}%
\providecommand \bibitemNoStop [0]{.\EOS\space}%
\providecommand \EOS [0]{\spacefactor3000\relax}%
\providecommand \BibitemShut  [1]{\csname bibitem#1\endcsname}%
\let\auto@bib@innerbib\@empty
\bibitem [{\citenamefont {Carney}\ \emph {et~al.}(2019)\citenamefont {Carney},
  \citenamefont {Stamp},\ and\ \citenamefont {Taylor}}]{Carney:2018ofe}%
  \BibitemOpen
  \bibfield  {author} {\bibinfo {author} {\bibfnamefont {D.}~\bibnamefont
  {Carney}}, \bibinfo {author} {\bibfnamefont {P.~C.~E.}\ \bibnamefont
  {Stamp}}, \ and\ \bibinfo {author} {\bibfnamefont {J.~M.}\ \bibnamefont
  {Taylor}},\ }\bibfield  {title} {\enquote {\bibinfo {title} {{Tabletop
  experiments for quantum gravity: a user\textquoteright{}s manual}},}\ }\href
  {\doibase 10.1088/1361-6382/aaf9ca} {\bibfield  {journal} {\bibinfo
  {journal} {Class. Quant. Grav.}\ }\textbf {\bibinfo {volume} {36}},\ \bibinfo
  {pages} {034001} (\bibinfo {year} {2019})},\ \Eprint
  {http://arxiv.org/abs/1807.11494} {arXiv:1807.11494 [quant-ph]} \BibitemShut
  {NoStop}%
\bibitem [{\citenamefont {Kafri}\ and\ \citenamefont
  {Taylor}(2013)}]{kafri2013noise}%
  \BibitemOpen
  \bibfield  {author} {\bibinfo {author} {\bibfnamefont {D.}~\bibnamefont
  {Kafri}}\ and\ \bibinfo {author} {\bibfnamefont {J.}~\bibnamefont {Taylor}},\
  }\bibfield  {title} {\enquote {\bibinfo {title} {A noise inequality for
  classical forces},}\ }\href@noop {} {\  (\bibinfo {year} {2013})},\ \Eprint
  {http://arxiv.org/abs/1311.4558} {arXiv:1311.4558 [quant-ph]} \BibitemShut
  {NoStop}%
\bibitem [{\citenamefont {Kafri}\ \emph {et~al.}(2014)\citenamefont {Kafri},
  \citenamefont {Taylor},\ and\ \citenamefont {Milburn}}]{Kafri:2014zsa}%
  \BibitemOpen
  \bibfield  {author} {\bibinfo {author} {\bibfnamefont {D.}~\bibnamefont
  {Kafri}}, \bibinfo {author} {\bibfnamefont {J.~M.}\ \bibnamefont {Taylor}}, \
  and\ \bibinfo {author} {\bibfnamefont {G.~J.}\ \bibnamefont {Milburn}},\
  }\bibfield  {title} {\enquote {\bibinfo {title} {{A classical channel model
  for gravitational decoherence}},}\ }\href {\doibase
  10.1088/1367-2630/16/6/065020} {\bibfield  {journal} {\bibinfo  {journal}
  {New J. Phys.}\ }\textbf {\bibinfo {volume} {16}},\ \bibinfo {pages} {065020}
  (\bibinfo {year} {2014})},\ \Eprint {http://arxiv.org/abs/1401.0946}
  {arXiv:1401.0946 [quant-ph]} \BibitemShut {NoStop}%
\bibitem [{\citenamefont {Kafri}\ \emph {et~al.}(2015)\citenamefont {Kafri},
  \citenamefont {Milburn},\ and\ \citenamefont {Taylor}}]{kafri2015bounds}%
  \BibitemOpen
  \bibfield  {author} {\bibinfo {author} {\bibfnamefont {D.}~\bibnamefont
  {Kafri}}, \bibinfo {author} {\bibfnamefont {G.}~\bibnamefont {Milburn}}, \
  and\ \bibinfo {author} {\bibfnamefont {J.}~\bibnamefont {Taylor}},\
  }\bibfield  {title} {\enquote {\bibinfo {title} {Bounds on quantum
  communication via newtonian gravity},}\ }\href@noop {} {\bibfield  {journal}
  {\bibinfo  {journal} {New Journal of Physics}\ }\textbf {\bibinfo {volume}
  {17}},\ \bibinfo {pages} {015006} (\bibinfo {year} {2015})}\BibitemShut
  {NoStop}%
\bibitem [{\citenamefont {Bahrami}\ \emph {et~al.}(2015)\citenamefont
  {Bahrami}, \citenamefont {Bassi}, \citenamefont {McMillen}, \citenamefont
  {Paternostro},\ and\ \citenamefont {Ulbricht}}]{Bahrami:2015wma}%
  \BibitemOpen
  \bibfield  {author} {\bibinfo {author} {\bibfnamefont {M.}~\bibnamefont
  {Bahrami}}, \bibinfo {author} {\bibfnamefont {A.}~\bibnamefont {Bassi}},
  \bibinfo {author} {\bibfnamefont {S.}~\bibnamefont {McMillen}}, \bibinfo
  {author} {\bibfnamefont {M.}~\bibnamefont {Paternostro}}, \ and\ \bibinfo
  {author} {\bibfnamefont {H.}~\bibnamefont {Ulbricht}},\ }\bibfield  {title}
  {\enquote {\bibinfo {title} {{Is Gravity Quantum?}}}\ }\href@noop {} {\
  (\bibinfo {year} {2015})},\ \Eprint {http://arxiv.org/abs/1507.05733}
  {arXiv:1507.05733 [quant-ph]} \BibitemShut {NoStop}%
\bibitem [{\citenamefont {Anastopoulos}\ and\ \citenamefont
  {Hu}(2015)}]{anastopoulos2015probing}%
  \BibitemOpen
  \bibfield  {author} {\bibinfo {author} {\bibfnamefont {C.}~\bibnamefont
  {Anastopoulos}}\ and\ \bibinfo {author} {\bibfnamefont {B.-L.}\ \bibnamefont
  {Hu}},\ }\bibfield  {title} {\enquote {\bibinfo {title} {Probing a
  gravitational cat state},}\ }\href@noop {} {\bibfield  {journal} {\bibinfo
  {journal} {Class. Quantum Gravity}\ }\textbf {\bibinfo {volume} {32}},\
  \bibinfo {pages} {165022} (\bibinfo {year} {2015})}\BibitemShut {NoStop}%
\bibitem [{\citenamefont {Bose}\ \emph {et~al.}(2017)\citenamefont {Bose},
  \citenamefont {Mazumdar}, \citenamefont {Morley}, \citenamefont {Ulbricht},
  \citenamefont {Toro{\v{s}}}, \citenamefont {Paternostro}, \citenamefont
  {Geraci}, \citenamefont {Barker}, \citenamefont {Kim},\ and\ \citenamefont
  {Milburn}}]{bose2017spin}%
  \BibitemOpen
  \bibfield  {author} {\bibinfo {author} {\bibfnamefont {S.}~\bibnamefont
  {Bose}}, \bibinfo {author} {\bibfnamefont {A.}~\bibnamefont {Mazumdar}},
  \bibinfo {author} {\bibfnamefont {G.~W.}\ \bibnamefont {Morley}}, \bibinfo
  {author} {\bibfnamefont {H.}~\bibnamefont {Ulbricht}}, \bibinfo {author}
  {\bibfnamefont {M.}~\bibnamefont {Toro{\v{s}}}}, \bibinfo {author}
  {\bibfnamefont {M.}~\bibnamefont {Paternostro}}, \bibinfo {author}
  {\bibfnamefont {A.~A.}\ \bibnamefont {Geraci}}, \bibinfo {author}
  {\bibfnamefont {P.~F.}\ \bibnamefont {Barker}}, \bibinfo {author}
  {\bibfnamefont {M.}~\bibnamefont {Kim}}, \ and\ \bibinfo {author}
  {\bibfnamefont {G.}~\bibnamefont {Milburn}},\ }\bibfield  {title} {\enquote
  {\bibinfo {title} {Spin entanglement witness for quantum gravity},}\
  }\href@noop {} {\bibfield  {journal} {\bibinfo  {journal} {Phys. Rev. Lett.}\
  }\textbf {\bibinfo {volume} {119}},\ \bibinfo {pages} {240401} (\bibinfo
  {year} {2017})}\BibitemShut {NoStop}%
\bibitem [{\citenamefont {Marletto}\ and\ \citenamefont
  {Vedral}(2017)}]{marletto2017gravitationally}%
  \BibitemOpen
  \bibfield  {author} {\bibinfo {author} {\bibfnamefont {C.}~\bibnamefont
  {Marletto}}\ and\ \bibinfo {author} {\bibfnamefont {V.}~\bibnamefont
  {Vedral}},\ }\bibfield  {title} {\enquote {\bibinfo {title} {Gravitationally
  induced entanglement between two massive particles is sufficient evidence of
  quantum effects in gravity},}\ }\href@noop {} {\bibfield  {journal} {\bibinfo
   {journal} {Phys. Rev. Lett.}\ }\textbf {\bibinfo {volume} {119}},\ \bibinfo
  {pages} {240402} (\bibinfo {year} {2017})}\BibitemShut {NoStop}%
\bibitem [{\citenamefont {Haine}(2021)}]{haine2021searching}%
  \BibitemOpen
  \bibfield  {author} {\bibinfo {author} {\bibfnamefont {S.~A.}\ \bibnamefont
  {Haine}},\ }\bibfield  {title} {\enquote {\bibinfo {title} {Searching for
  signatures of quantum gravity in quantum gases},}\ }\href@noop {} {\bibfield
  {journal} {\bibinfo  {journal} {New J. Phys.}\ }\textbf {\bibinfo {volume}
  {23}},\ \bibinfo {pages} {033020} (\bibinfo {year} {2021})}\BibitemShut
  {NoStop}%
\bibitem [{\citenamefont {Qvarfort}\ \emph {et~al.}(2020)\citenamefont
  {Qvarfort}, \citenamefont {Bose},\ and\ \citenamefont
  {Serafini}}]{Qvarfort:2018uag}%
  \BibitemOpen
  \bibfield  {author} {\bibinfo {author} {\bibfnamefont {S.}~\bibnamefont
  {Qvarfort}}, \bibinfo {author} {\bibfnamefont {S.}~\bibnamefont {Bose}}, \
  and\ \bibinfo {author} {\bibfnamefont {A.}~\bibnamefont {Serafini}},\
  }\bibfield  {title} {\enquote {\bibinfo {title} {{Mesoscopic entanglement
  through central\textendash{}potential interactions}},}\ }\href {\doibase
  10.1088/1361-6455/abbe8d} {\bibfield  {journal} {\bibinfo  {journal} {J.
  Phys. B}\ }\textbf {\bibinfo {volume} {53}},\ \bibinfo {pages} {235501}
  (\bibinfo {year} {2020})},\ \Eprint {http://arxiv.org/abs/1812.09776}
  {arXiv:1812.09776 [quant-ph]} \BibitemShut {NoStop}%
\bibitem [{\citenamefont {Carlesso}\ \emph {et~al.}(2019)\citenamefont
  {Carlesso}, \citenamefont {Bassi}, \citenamefont {Paternostro},\ and\
  \citenamefont {Ulbricht}}]{Carlesso:2019cuh}%
  \BibitemOpen
  \bibfield  {author} {\bibinfo {author} {\bibfnamefont {M.}~\bibnamefont
  {Carlesso}}, \bibinfo {author} {\bibfnamefont {A.}~\bibnamefont {Bassi}},
  \bibinfo {author} {\bibfnamefont {M.}~\bibnamefont {Paternostro}}, \ and\
  \bibinfo {author} {\bibfnamefont {H.}~\bibnamefont {Ulbricht}},\ }\bibfield
  {title} {\enquote {\bibinfo {title} {{Testing the gravitational field
  generated by a quantum superposition}},}\ }\href {\doibase
  10.1088/1367-2630/ab41c1} {\bibfield  {journal} {\bibinfo  {journal} {New J.
  Phys.}\ }\textbf {\bibinfo {volume} {21}},\ \bibinfo {pages} {093052}
  (\bibinfo {year} {2019})},\ \Eprint {http://arxiv.org/abs/1906.04513}
  {arXiv:1906.04513 [quant-ph]} \BibitemShut {NoStop}%
\bibitem [{\citenamefont {Miao}\ \emph {et~al.}(2020)\citenamefont {Miao},
  \citenamefont {Martynov}, \citenamefont {Yang},\ and\ \citenamefont
  {Datta}}]{PhysRevA.101.063804}%
  \BibitemOpen
  \bibfield  {author} {\bibinfo {author} {\bibfnamefont {H.}~\bibnamefont
  {Miao}}, \bibinfo {author} {\bibfnamefont {D.}~\bibnamefont {Martynov}},
  \bibinfo {author} {\bibfnamefont {H.}~\bibnamefont {Yang}}, \ and\ \bibinfo
  {author} {\bibfnamefont {A.}~\bibnamefont {Datta}},\ }\bibfield  {title}
  {\enquote {\bibinfo {title} {Quantum correlations of light mediated by
  gravity},}\ }\href {\doibase 10.1103/PhysRevA.101.063804} {\bibfield
  {journal} {\bibinfo  {journal} {Phys. Rev. A}\ }\textbf {\bibinfo {volume}
  {101}},\ \bibinfo {pages} {063804} (\bibinfo {year} {2020})}\BibitemShut
  {NoStop}%
\bibitem [{\citenamefont {Howl}\ \emph {et~al.}(2021)\citenamefont {Howl},
  \citenamefont {Vedral}, \citenamefont {Naik}, \citenamefont {Christodoulou},
  \citenamefont {Rovelli},\ and\ \citenamefont {Iyer}}]{howl2021non}%
  \BibitemOpen
  \bibfield  {author} {\bibinfo {author} {\bibfnamefont {R.}~\bibnamefont
  {Howl}}, \bibinfo {author} {\bibfnamefont {V.}~\bibnamefont {Vedral}},
  \bibinfo {author} {\bibfnamefont {D.}~\bibnamefont {Naik}}, \bibinfo {author}
  {\bibfnamefont {M.}~\bibnamefont {Christodoulou}}, \bibinfo {author}
  {\bibfnamefont {C.}~\bibnamefont {Rovelli}}, \ and\ \bibinfo {author}
  {\bibfnamefont {A.}~\bibnamefont {Iyer}},\ }\bibfield  {title} {\enquote
  {\bibinfo {title} {Non-gaussianity as a signature of a quantum theory of
  gravity},}\ }\href@noop {} {\bibfield  {journal} {\bibinfo  {journal} {PRX
  Quantum}\ }\textbf {\bibinfo {volume} {2}},\ \bibinfo {pages} {010325}
  (\bibinfo {year} {2021})}\BibitemShut {NoStop}%
\bibitem [{\citenamefont {Matsumura}\ and\ \citenamefont
  {Yamamoto}(2020)}]{Matsumura:2020law}%
  \BibitemOpen
  \bibfield  {author} {\bibinfo {author} {\bibfnamefont {A.}~\bibnamefont
  {Matsumura}}\ and\ \bibinfo {author} {\bibfnamefont {K.}~\bibnamefont
  {Yamamoto}},\ }\bibfield  {title} {\enquote {\bibinfo {title}
  {{Gravity-induced entanglement in optomechanical systems}},}\ }\href
  {\doibase 10.1103/PhysRevD.102.106021} {\bibfield  {journal} {\bibinfo
  {journal} {Phys. Rev. D}\ }\textbf {\bibinfo {volume} {102}},\ \bibinfo
  {pages} {106021} (\bibinfo {year} {2020})},\ \Eprint
  {http://arxiv.org/abs/2010.05161} {arXiv:2010.05161 [gr-qc]} \BibitemShut
  {NoStop}%
\bibitem [{\citenamefont {Pedernales}\ \emph {et~al.}(2022)\citenamefont
  {Pedernales}, \citenamefont {Streltsov},\ and\ \citenamefont
  {Plenio}}]{Pedernales:2021dja}%
  \BibitemOpen
  \bibfield  {author} {\bibinfo {author} {\bibfnamefont {J.~S.}\ \bibnamefont
  {Pedernales}}, \bibinfo {author} {\bibfnamefont {K.}~\bibnamefont
  {Streltsov}}, \ and\ \bibinfo {author} {\bibfnamefont {M.~B.}\ \bibnamefont
  {Plenio}},\ }\bibfield  {title} {\enquote {\bibinfo {title} {{Enhancing
  Gravitational Interaction between Quantum Systems by a Massive Mediator}},}\
  }\href {\doibase 10.1103/PhysRevLett.128.110401} {\bibfield  {journal}
  {\bibinfo  {journal} {Phys. Rev. Lett.}\ }\textbf {\bibinfo {volume} {128}},\
  \bibinfo {pages} {110401} (\bibinfo {year} {2022})},\ \Eprint
  {http://arxiv.org/abs/2104.14524} {arXiv:2104.14524 [quant-ph]} \BibitemShut
  {NoStop}%
\bibitem [{\citenamefont {Liu}\ \emph {et~al.}(2021)\citenamefont {Liu},
  \citenamefont {Mummery}, \citenamefont {Zhou},\ and\ \citenamefont
  {Sillanp{\"a}{\"a}}}]{liu2021gravitational}%
  \BibitemOpen
  \bibfield  {author} {\bibinfo {author} {\bibfnamefont {Y.}~\bibnamefont
  {Liu}}, \bibinfo {author} {\bibfnamefont {J.}~\bibnamefont {Mummery}},
  \bibinfo {author} {\bibfnamefont {J.}~\bibnamefont {Zhou}}, \ and\ \bibinfo
  {author} {\bibfnamefont {M.~A.}\ \bibnamefont {Sillanp{\"a}{\"a}}},\
  }\bibfield  {title} {\enquote {\bibinfo {title} {Gravitational forces between
  nonclassical mechanical oscillators},}\ }\href@noop {} {\bibfield  {journal}
  {\bibinfo  {journal} {Phys. Rev. Applied}\ }\textbf {\bibinfo {volume}
  {15}},\ \bibinfo {pages} {034004} (\bibinfo {year} {2021})}\BibitemShut
  {NoStop}%
\bibitem [{\citenamefont {Datta}\ and\ \citenamefont
  {Miao}(2021)}]{Datta:2021ywm}%
  \BibitemOpen
  \bibfield  {author} {\bibinfo {author} {\bibfnamefont {A.}~\bibnamefont
  {Datta}}\ and\ \bibinfo {author} {\bibfnamefont {H.}~\bibnamefont {Miao}},\
  }\bibfield  {title} {\enquote {\bibinfo {title} {{Signatures of the quantum
  nature of gravity in the differential motion of two masses}},}\ }\href@noop
  {} {\  (\bibinfo {year} {2021})},\ \Eprint {http://arxiv.org/abs/2104.04414}
  {arXiv:2104.04414 [gr-qc]} \BibitemShut {NoStop}%
\bibitem [{\citenamefont {Feynman}(1957)}]{cecile2011role}%
  \BibitemOpen
  \bibfield  {author} {\bibinfo {author} {\bibfnamefont {R.}~\bibnamefont
  {Feynman}},\ }\bibfield  {title} {\enquote {\bibinfo {title} {The role of
  gravitation in physics: report from the 1957 chapel hill conference},}\
  }\href@noop {} {\  (\bibinfo {year} {1957})}\BibitemShut {NoStop}%
\bibitem [{\citenamefont {Feynman}(1971)}]{feynman1971lectures}%
  \BibitemOpen
  \bibfield  {author} {\bibinfo {author} {\bibfnamefont {R.}~\bibnamefont
  {Feynman}},\ }\href@noop {} {\emph {\bibinfo {title} {Lectures on
  Gravitation, 1962-63}}}\ (\bibinfo  {publisher} {California Institute of
  Technology Press},\ \bibinfo {year} {1971})\BibitemShut {NoStop}%
\bibitem [{\citenamefont {Page}\ and\ \citenamefont
  {Geilker}(1981)}]{page1981indirect}%
  \BibitemOpen
  \bibfield  {author} {\bibinfo {author} {\bibfnamefont {D.~N.}\ \bibnamefont
  {Page}}\ and\ \bibinfo {author} {\bibfnamefont {C.}~\bibnamefont {Geilker}},\
  }\bibfield  {title} {\enquote {\bibinfo {title} {Indirect evidence for
  quantum gravity},}\ }\href@noop {} {\bibfield  {journal} {\bibinfo  {journal}
  {Phys. Rev. Lett.}\ }\textbf {\bibinfo {volume} {47}},\ \bibinfo {pages}
  {979} (\bibinfo {year} {1981})}\BibitemShut {NoStop}%
\bibitem [{\citenamefont {Donoghue}(1994)}]{Donoghue:1994dn}%
  \BibitemOpen
  \bibfield  {author} {\bibinfo {author} {\bibfnamefont {J.~F.}\ \bibnamefont
  {Donoghue}},\ }\bibfield  {title} {\enquote {\bibinfo {title} {{General
  relativity as an effective field theory: The leading quantum corrections}},}\
  }\href {\doibase 10.1103/PhysRevD.50.3874} {\bibfield  {journal} {\bibinfo
  {journal} {Phys. Rev.}\ }\textbf {\bibinfo {volume} {D50}},\ \bibinfo {pages}
  {3874--3888} (\bibinfo {year} {1994})},\ \Eprint
  {http://arxiv.org/abs/gr-qc/9405057} {arXiv:gr-qc/9405057 [gr-qc]}
  \BibitemShut {NoStop}%
\bibitem [{\citenamefont {Donoghue}(1995)}]{Donoghue:1995cz}%
  \BibitemOpen
  \bibfield  {author} {\bibinfo {author} {\bibfnamefont {J.~F.}\ \bibnamefont
  {Donoghue}},\ }\bibfield  {title} {\enquote {\bibinfo {title} {{Introduction
  to the effective field theory description of gravity}},}\ }in\ \href@noop {}
  {\emph {\bibinfo {booktitle} {{Advanced School on Effective Theories
  Almunecar, Spain, June 25-July 1, 1995}}}}\ (\bibinfo {year} {1995})\ \Eprint
  {http://arxiv.org/abs/gr-qc/9512024} {arXiv:gr-qc/9512024 [gr-qc]}
  \BibitemShut {NoStop}%
\bibitem [{\citenamefont {Burgess}(2004)}]{Burgess:2003jk}%
  \BibitemOpen
  \bibfield  {author} {\bibinfo {author} {\bibfnamefont {C.~P.}\ \bibnamefont
  {Burgess}},\ }\bibfield  {title} {\enquote {\bibinfo {title} {{Quantum
  gravity in everyday life: General relativity as an effective field
  theory}},}\ }\href {\doibase 10.12942/lrr-2004-5} {\bibfield  {journal}
  {\bibinfo  {journal} {Living Rev. Rel.}\ }\textbf {\bibinfo {volume} {7}},\
  \bibinfo {pages} {5--56} (\bibinfo {year} {2004})},\ \Eprint
  {http://arxiv.org/abs/gr-qc/0311082} {arXiv:gr-qc/0311082 [gr-qc]}
  \BibitemShut {NoStop}%
\bibitem [{\citenamefont {Carney}(2022)}]{Carney:2021vvt}%
  \BibitemOpen
  \bibfield  {author} {\bibinfo {author} {\bibfnamefont {D.}~\bibnamefont
  {Carney}},\ }\bibfield  {title} {\enquote {\bibinfo {title} {{Newton,
  entanglement, and the graviton}},}\ }\href {\doibase
  10.1103/PhysRevD.105.024029} {\bibfield  {journal} {\bibinfo  {journal}
  {Phys. Rev. D}\ }\textbf {\bibinfo {volume} {105}},\ \bibinfo {pages}
  {024029} (\bibinfo {year} {2022})},\ \Eprint
  {http://arxiv.org/abs/2108.06320} {arXiv:2108.06320 [quant-ph]} \BibitemShut
  {NoStop}%
\bibitem [{\citenamefont {Tilloy}\ and\ \citenamefont
  {Di{\'o}si}(2016)}]{tilloy2016sourcing}%
  \BibitemOpen
  \bibfield  {author} {\bibinfo {author} {\bibfnamefont {A.}~\bibnamefont
  {Tilloy}}\ and\ \bibinfo {author} {\bibfnamefont {L.}~\bibnamefont
  {Di{\'o}si}},\ }\bibfield  {title} {\enquote {\bibinfo {title} {Sourcing
  semiclassical gravity from spontaneously localized quantum matter},}\
  }\href@noop {} {\bibfield  {journal} {\bibinfo  {journal} {Physical Review
  D}\ }\textbf {\bibinfo {volume} {93}},\ \bibinfo {pages} {024026} (\bibinfo
  {year} {2016})}\BibitemShut {NoStop}%
\bibitem [{\citenamefont {Hall}\ and\ \citenamefont
  {Reginatto}(2018)}]{Hall:2017nzl}%
  \BibitemOpen
  \bibfield  {author} {\bibinfo {author} {\bibfnamefont {M.~J.~W.}\
  \bibnamefont {Hall}}\ and\ \bibinfo {author} {\bibfnamefont {M.}~\bibnamefont
  {Reginatto}},\ }\bibfield  {title} {\enquote {\bibinfo {title} {{On two
  recent proposals for witnessing nonclassical gravity}},}\ }\href {\doibase
  10.1088/1751-8121/aaa734} {\bibfield  {journal} {\bibinfo  {journal} {J.
  Phys. A}\ }\textbf {\bibinfo {volume} {51}},\ \bibinfo {pages} {085303}
  (\bibinfo {year} {2018})},\ \Eprint {http://arxiv.org/abs/1707.07974}
  {arXiv:1707.07974 [quant-ph]} \BibitemShut {NoStop}%
\bibitem [{\citenamefont {Tilloy}(2018)}]{Tilloy:2018tjp}%
  \BibitemOpen
  \bibfield  {author} {\bibinfo {author} {\bibfnamefont {A.}~\bibnamefont
  {Tilloy}},\ }\bibfield  {title} {\enquote {\bibinfo {title} {{Binding quantum
  matter and space-time, without romanticism}},}\ }\href {\doibase
  10.1007/s10701-018-0224-6} {\bibfield  {journal} {\bibinfo  {journal} {Found.
  Phys.}\ }\textbf {\bibinfo {volume} {48}},\ \bibinfo {pages} {1753--1769}
  (\bibinfo {year} {2018})},\ \Eprint {http://arxiv.org/abs/1802.03291}
  {arXiv:1802.03291 [physics.hist-ph]} \BibitemShut {NoStop}%
\bibitem [{\citenamefont {Oppenheim}(2018)}]{Oppenheim:2018igd}%
  \BibitemOpen
  \bibfield  {author} {\bibinfo {author} {\bibfnamefont {J.}~\bibnamefont
  {Oppenheim}},\ }\bibfield  {title} {\enquote {\bibinfo {title} {{A
  post-quantum theory of classical gravity?}}}\ }\href@noop {} {\  (\bibinfo
  {year} {2018})},\ \Eprint {http://arxiv.org/abs/1811.03116} {arXiv:1811.03116
  [hep-th]} \BibitemShut {NoStop}%
\bibitem [{\citenamefont {Gro\ss{}ardt}(2022)}]{Grossardt:2022zsi}%
  \BibitemOpen
  \bibfield  {author} {\bibinfo {author} {\bibfnamefont {A.}~\bibnamefont
  {Gro\ss{}ardt}},\ }\bibfield  {title} {\enquote {\bibinfo {title} {{Three
  little paradoxes: Making sense of semiclassical gravity}},}\ }\href {\doibase
  10.1116/5.0073509} {\bibfield  {journal} {\bibinfo  {journal} {AVS Quantum
  Sci.}\ }\textbf {\bibinfo {volume} {4}},\ \bibinfo {pages} {010502} (\bibinfo
  {year} {2022})},\ \Eprint {http://arxiv.org/abs/2201.10452} {arXiv:2201.10452
  [gr-qc]} \BibitemShut {NoStop}%
\bibitem [{\citenamefont {Giulini}\ \emph {et~al.}(2022)\citenamefont
  {Giulini}, \citenamefont {Gro\ss{}ardt},\ and\ \citenamefont
  {Schwartz}}]{Giulini:2022wyp}%
  \BibitemOpen
  \bibfield  {author} {\bibinfo {author} {\bibfnamefont {D.}~\bibnamefont
  {Giulini}}, \bibinfo {author} {\bibfnamefont {A.}~\bibnamefont
  {Gro\ss{}ardt}}, \ and\ \bibinfo {author} {\bibfnamefont {P.~K.}\
  \bibnamefont {Schwartz}},\ }\bibfield  {title} {\enquote {\bibinfo {title}
  {{Coupling Quantum Matter and Gravity}},}\ }\href@noop {} {\  (\bibinfo
  {year} {2022})},\ \Eprint {http://arxiv.org/abs/2207.05029} {arXiv:2207.05029
  [gr-qc]} \BibitemShut {NoStop}%
\bibitem [{\citenamefont {Layton}\ \emph {et~al.}(2022)\citenamefont {Layton},
  \citenamefont {Oppenheim},\ and\ \citenamefont
  {Weller-Davies}}]{Layton:2022sku}%
  \BibitemOpen
  \bibfield  {author} {\bibinfo {author} {\bibfnamefont {I.}~\bibnamefont
  {Layton}}, \bibinfo {author} {\bibfnamefont {J.}~\bibnamefont {Oppenheim}}, \
  and\ \bibinfo {author} {\bibfnamefont {Z.}~\bibnamefont {Weller-Davies}},\
  }\bibfield  {title} {\enquote {\bibinfo {title} {{A healthier semi-classical
  dynamics}},}\ }\href@noop {} {\  (\bibinfo {year} {2022})},\ \Eprint
  {http://arxiv.org/abs/2208.11722} {arXiv:2208.11722 [quant-ph]} \BibitemShut
  {NoStop}%
\bibitem [{\citenamefont {Ma}\ \emph {et~al.}(2022)\citenamefont {Ma},
  \citenamefont {Guff}, \citenamefont {Morley}, \citenamefont {Pikovski},\ and\
  \citenamefont {Kim}}]{ma2022limits}%
  \BibitemOpen
  \bibfield  {author} {\bibinfo {author} {\bibfnamefont {Y.}~\bibnamefont
  {Ma}}, \bibinfo {author} {\bibfnamefont {T.}~\bibnamefont {Guff}}, \bibinfo
  {author} {\bibfnamefont {G.~W.}\ \bibnamefont {Morley}}, \bibinfo {author}
  {\bibfnamefont {I.}~\bibnamefont {Pikovski}}, \ and\ \bibinfo {author}
  {\bibfnamefont {M.}~\bibnamefont {Kim}},\ }\bibfield  {title} {\enquote
  {\bibinfo {title} {Limits on inference of gravitational entanglement},}\
  }\href@noop {} {\bibfield  {journal} {\bibinfo  {journal} {Physical Review
  Research}\ }\textbf {\bibinfo {volume} {4}},\ \bibinfo {pages} {013024}
  (\bibinfo {year} {2022})}\BibitemShut {NoStop}%
\bibitem [{\citenamefont {Polchinski}(1991)}]{polchinski1991weinberg}%
  \BibitemOpen
  \bibfield  {author} {\bibinfo {author} {\bibfnamefont {J.}~\bibnamefont
  {Polchinski}},\ }\bibfield  {title} {\enquote {\bibinfo {title} {Weinberg’s
  nonlinear quantum mechanics and the einstein-podolsky-rosen paradox},}\
  }\href@noop {} {\bibfield  {journal} {\bibinfo  {journal} {Physical Review
  Letters}\ }\textbf {\bibinfo {volume} {66}},\ \bibinfo {pages} {397}
  (\bibinfo {year} {1991})}\BibitemShut {NoStop}%
\bibitem [{\citenamefont {Banks}\ \emph {et~al.}(1984)\citenamefont {Banks},
  \citenamefont {Susskind},\ and\ \citenamefont
  {Peskin}}]{banks1984difficulties}%
  \BibitemOpen
  \bibfield  {author} {\bibinfo {author} {\bibfnamefont {T.}~\bibnamefont
  {Banks}}, \bibinfo {author} {\bibfnamefont {L.}~\bibnamefont {Susskind}}, \
  and\ \bibinfo {author} {\bibfnamefont {M.~E.}\ \bibnamefont {Peskin}},\
  }\bibfield  {title} {\enquote {\bibinfo {title} {Difficulties for the
  evolution of pure states into mixed states},}\ }\href@noop {} {\bibfield
  {journal} {\bibinfo  {journal} {Nuclear Physics B}\ }\textbf {\bibinfo
  {volume} {244}},\ \bibinfo {pages} {125--134} (\bibinfo {year}
  {1984})}\BibitemShut {NoStop}%
\bibitem [{\citenamefont {Bassi}\ \emph {et~al.}(2013)\citenamefont {Bassi},
  \citenamefont {Lochan}, \citenamefont {Satin}, \citenamefont {Singh},\ and\
  \citenamefont {Ulbricht}}]{bassi2013models}%
  \BibitemOpen
  \bibfield  {author} {\bibinfo {author} {\bibfnamefont {A.}~\bibnamefont
  {Bassi}}, \bibinfo {author} {\bibfnamefont {K.}~\bibnamefont {Lochan}},
  \bibinfo {author} {\bibfnamefont {S.}~\bibnamefont {Satin}}, \bibinfo
  {author} {\bibfnamefont {T.~P.}\ \bibnamefont {Singh}}, \ and\ \bibinfo
  {author} {\bibfnamefont {H.}~\bibnamefont {Ulbricht}},\ }\bibfield  {title}
  {\enquote {\bibinfo {title} {Models of wave-function collapse, underlying
  theories, and experimental tests},}\ }\href@noop {} {\bibfield  {journal}
  {\bibinfo  {journal} {Reviews of Modern Physics}\ }\textbf {\bibinfo {volume}
  {85}},\ \bibinfo {pages} {471} (\bibinfo {year} {2013})}\BibitemShut
  {NoStop}%
\bibitem [{\citenamefont {Carney}\ \emph {et~al.}(2021)\citenamefont {Carney},
  \citenamefont {M\"uller},\ and\ \citenamefont {Taylor}}]{Carney:2021yfw}%
  \BibitemOpen
  \bibfield  {author} {\bibinfo {author} {\bibfnamefont {D.}~\bibnamefont
  {Carney}}, \bibinfo {author} {\bibfnamefont {H.}~\bibnamefont {M\"uller}}, \
  and\ \bibinfo {author} {\bibfnamefont {J.~M.}\ \bibnamefont {Taylor}},\
  }\bibfield  {title} {\enquote {\bibinfo {title} {{Using an Atom
  Interferometer to Infer Gravitational Entanglement Generation}},}\ }\href
  {\doibase 10.1103/PRXQuantum.2.030330} {\bibfield  {journal} {\bibinfo
  {journal} {PRX Quantum}\ }\textbf {\bibinfo {volume} {2}},\ \bibinfo {pages}
  {030330} (\bibinfo {year} {2021})},\ \bibinfo {note} {[Erratum: PRX Quantum
  3, 010902 (2022)]},\ \Eprint {http://arxiv.org/abs/2101.11629}
  {arXiv:2101.11629 [quant-ph]} \BibitemShut {NoStop}%
\bibitem [{\citenamefont {Streltsov}\ \emph {et~al.}(2022)\citenamefont
  {Streltsov}, \citenamefont {Pedernales},\ and\ \citenamefont
  {Plenio}}]{Streltsov:2021ahn}%
  \BibitemOpen
  \bibfield  {author} {\bibinfo {author} {\bibfnamefont {K.}~\bibnamefont
  {Streltsov}}, \bibinfo {author} {\bibfnamefont {J.~S.}\ \bibnamefont
  {Pedernales}}, \ and\ \bibinfo {author} {\bibfnamefont {M.~B.}\ \bibnamefont
  {Plenio}},\ }\bibfield  {title} {\enquote {\bibinfo {title} {{On the
  Significance of Interferometric Revivals for the Fundamental Description of
  Gravity}},}\ }\href {\doibase 10.3390/universe8020058} {\bibfield  {journal}
  {\bibinfo  {journal} {Universe}\ }\textbf {\bibinfo {volume} {8}},\ \bibinfo
  {pages} {58} (\bibinfo {year} {2022})},\ \Eprint
  {http://arxiv.org/abs/2111.04570} {arXiv:2111.04570 [quant-ph]} \BibitemShut
  {NoStop}%
\bibitem [{\citenamefont {Hamilton}\ \emph {et~al.}(2015)\citenamefont
  {Hamilton}, \citenamefont {Jaffe}, \citenamefont {Haslinger}, \citenamefont
  {Simmons}, \citenamefont {M{\"u}ller},\ and\ \citenamefont
  {Khoury}}]{hamilton2015atom}%
  \BibitemOpen
  \bibfield  {author} {\bibinfo {author} {\bibfnamefont {P.}~\bibnamefont
  {Hamilton}}, \bibinfo {author} {\bibfnamefont {M.}~\bibnamefont {Jaffe}},
  \bibinfo {author} {\bibfnamefont {P.}~\bibnamefont {Haslinger}}, \bibinfo
  {author} {\bibfnamefont {Q.}~\bibnamefont {Simmons}}, \bibinfo {author}
  {\bibfnamefont {H.}~\bibnamefont {M{\"u}ller}}, \ and\ \bibinfo {author}
  {\bibfnamefont {J.}~\bibnamefont {Khoury}},\ }\bibfield  {title} {\enquote
  {\bibinfo {title} {Atom-interferometry constraints on dark energy},}\
  }\href@noop {} {\bibfield  {journal} {\bibinfo  {journal} {Science}\ }\textbf
  {\bibinfo {volume} {349}},\ \bibinfo {pages} {849--851} (\bibinfo {year}
  {2015})}\BibitemShut {NoStop}%
\bibitem [{\citenamefont {Xu}\ \emph {et~al.}(2019)\citenamefont {Xu},
  \citenamefont {Jaffe}, \citenamefont {Panda}, \citenamefont {Kristensen},
  \citenamefont {Clark},\ and\ \citenamefont {M{\"u}ller}}]{xu2019probing}%
  \BibitemOpen
  \bibfield  {author} {\bibinfo {author} {\bibfnamefont {V.}~\bibnamefont
  {Xu}}, \bibinfo {author} {\bibfnamefont {M.}~\bibnamefont {Jaffe}}, \bibinfo
  {author} {\bibfnamefont {C.~D.}\ \bibnamefont {Panda}}, \bibinfo {author}
  {\bibfnamefont {S.~L.}\ \bibnamefont {Kristensen}}, \bibinfo {author}
  {\bibfnamefont {L.~W.}\ \bibnamefont {Clark}}, \ and\ \bibinfo {author}
  {\bibfnamefont {H.}~\bibnamefont {M{\"u}ller}},\ }\bibfield  {title}
  {\enquote {\bibinfo {title} {Probing gravity by holding atoms for 20
  seconds},}\ }\href@noop {} {\bibfield  {journal} {\bibinfo  {journal}
  {Science}\ }\textbf {\bibinfo {volume} {366}},\ \bibinfo {pages} {745--749}
  (\bibinfo {year} {2019})}\BibitemShut {NoStop}%
\bibitem [{\citenamefont {Panda}\ \emph {et~al.}(2022)\citenamefont {Panda},
  \citenamefont {Tao}, \citenamefont {Egelhoff}, \citenamefont {Ceja},
  \citenamefont {Xu},\ and\ \citenamefont {M\"uller}}]{Panda:2022gtw}%
  \BibitemOpen
  \bibfield  {author} {\bibinfo {author} {\bibfnamefont {C.~D.}\ \bibnamefont
  {Panda}}, \bibinfo {author} {\bibfnamefont {M.}~\bibnamefont {Tao}}, \bibinfo
  {author} {\bibfnamefont {J.}~\bibnamefont {Egelhoff}}, \bibinfo {author}
  {\bibfnamefont {M.}~\bibnamefont {Ceja}}, \bibinfo {author} {\bibfnamefont
  {V.}~\bibnamefont {Xu}}, \ and\ \bibinfo {author} {\bibfnamefont
  {H.}~\bibnamefont {M\"uller}},\ }\bibfield  {title} {\enquote {\bibinfo
  {title} {{Quantum metrology by one-minute interrogation of a coherent atomic
  spatial superposition}},}\ }\href@noop {} {\  (\bibinfo {year} {2022})},\
  \Eprint {http://arxiv.org/abs/2210.07289} {arXiv:2210.07289
  [physics.atom-ph]} \BibitemShut {NoStop}%
\bibitem [{\citenamefont {Colella}\ \emph {et~al.}(1975)\citenamefont
  {Colella}, \citenamefont {Overhauser},\ and\ \citenamefont
  {Werner}}]{colella1975observation}%
  \BibitemOpen
  \bibfield  {author} {\bibinfo {author} {\bibfnamefont {R.}~\bibnamefont
  {Colella}}, \bibinfo {author} {\bibfnamefont {A.~W.}\ \bibnamefont
  {Overhauser}}, \ and\ \bibinfo {author} {\bibfnamefont {S.~A.}\ \bibnamefont
  {Werner}},\ }\bibfield  {title} {\enquote {\bibinfo {title} {Observation of
  gravitationally induced quantum interference},}\ }\href@noop {} {\bibfield
  {journal} {\bibinfo  {journal} {Physical Review Letters}\ }\textbf {\bibinfo
  {volume} {34}},\ \bibinfo {pages} {1472} (\bibinfo {year}
  {1975})}\BibitemShut {NoStop}%
\bibitem [{\citenamefont {Overstreet}\ \emph {et~al.}(2022)\citenamefont
  {Overstreet}, \citenamefont {Asenbaum}, \citenamefont {Curti}, \citenamefont
  {Kim},\ and\ \citenamefont {Kasevich}}]{overstreet2022observation}%
  \BibitemOpen
  \bibfield  {author} {\bibinfo {author} {\bibfnamefont {C.}~\bibnamefont
  {Overstreet}}, \bibinfo {author} {\bibfnamefont {P.}~\bibnamefont
  {Asenbaum}}, \bibinfo {author} {\bibfnamefont {J.}~\bibnamefont {Curti}},
  \bibinfo {author} {\bibfnamefont {M.}~\bibnamefont {Kim}}, \ and\ \bibinfo
  {author} {\bibfnamefont {M.~A.}\ \bibnamefont {Kasevich}},\ }\bibfield
  {title} {\enquote {\bibinfo {title} {Observation of a gravitational
  aharonov-bohm effect},}\ }\href@noop {} {\bibfield  {journal} {\bibinfo
  {journal} {Science}\ }\textbf {\bibinfo {volume} {375}},\ \bibinfo {pages}
  {226--229} (\bibinfo {year} {2022})}\BibitemShut {NoStop}%
\bibitem [{\citenamefont {Hosten}(2022)}]{Hosten:2021biw}%
  \BibitemOpen
  \bibfield  {author} {\bibinfo {author} {\bibfnamefont {O.}~\bibnamefont
  {Hosten}},\ }\bibfield  {title} {\enquote {\bibinfo {title} {{Constraints on
  probing quantum coherence to infer gravitational entanglement}},}\ }\href
  {\doibase 10.1103/PhysRevResearch.4.013023} {\bibfield  {journal} {\bibinfo
  {journal} {Phys. Rev. Res.}\ }\textbf {\bibinfo {volume} {4}},\ \bibinfo
  {pages} {013023} (\bibinfo {year} {2022})},\ \Eprint
  {http://arxiv.org/abs/2106.08221} {arXiv:2106.08221 [quant-ph]} \BibitemShut
  {NoStop}%
\bibitem [{\citenamefont {Zu}\ \emph {et~al.}(2022)\citenamefont {Zu},
  \citenamefont {Dai},\ and\ \citenamefont {de~Waele}}]{zu2022development}%
  \BibitemOpen
  \bibfield  {author} {\bibinfo {author} {\bibfnamefont {H.}~\bibnamefont
  {Zu}}, \bibinfo {author} {\bibfnamefont {W.}~\bibnamefont {Dai}}, \ and\
  \bibinfo {author} {\bibfnamefont {A.}~\bibnamefont {de~Waele}},\ }\bibfield
  {title} {\enquote {\bibinfo {title} {Development of dilution
  refrigerators—a review},}\ }\href@noop {} {\bibfield  {journal} {\bibinfo
  {journal} {Cryogenics}\ }\textbf {\bibinfo {volume} {121}},\ \bibinfo {pages}
  {103390} (\bibinfo {year} {2022})}\BibitemShut {NoStop}%
\bibitem [{\citenamefont {Scherschligt}\ \emph {et~al.}(2017)\citenamefont
  {Scherschligt}, \citenamefont {Fedchak}, \citenamefont {Barker},
  \citenamefont {Eckel}, \citenamefont {Klimov}, \citenamefont {Makrides},\
  and\ \citenamefont {Tiesinga}}]{scherschligt2017development}%
  \BibitemOpen
  \bibfield  {author} {\bibinfo {author} {\bibfnamefont {J.}~\bibnamefont
  {Scherschligt}}, \bibinfo {author} {\bibfnamefont {J.~A.}\ \bibnamefont
  {Fedchak}}, \bibinfo {author} {\bibfnamefont {D.~S.}\ \bibnamefont {Barker}},
  \bibinfo {author} {\bibfnamefont {S.}~\bibnamefont {Eckel}}, \bibinfo
  {author} {\bibfnamefont {N.}~\bibnamefont {Klimov}}, \bibinfo {author}
  {\bibfnamefont {C.}~\bibnamefont {Makrides}}, \ and\ \bibinfo {author}
  {\bibfnamefont {E.}~\bibnamefont {Tiesinga}},\ }\bibfield  {title} {\enquote
  {\bibinfo {title} {Development of a new uhv/xhv pressure standard (cold atom
  vacuum standard)},}\ }\href@noop {} {\bibfield  {journal} {\bibinfo
  {journal} {Metrologia}\ }\textbf {\bibinfo {volume} {54}},\ \bibinfo {pages}
  {S125} (\bibinfo {year} {2017})}\BibitemShut {NoStop}%
\bibitem [{\citenamefont {Gabrielse}\ \emph {et~al.}(1999)\citenamefont
  {Gabrielse}, \citenamefont {Khabbaz}, \citenamefont {Hall}, \citenamefont
  {Heimann}, \citenamefont {Kalinowsky},\ and\ \citenamefont
  {Jhe}}]{gabrielse1999precision}%
  \BibitemOpen
  \bibfield  {author} {\bibinfo {author} {\bibfnamefont {G.}~\bibnamefont
  {Gabrielse}}, \bibinfo {author} {\bibfnamefont {A.}~\bibnamefont {Khabbaz}},
  \bibinfo {author} {\bibfnamefont {D.}~\bibnamefont {Hall}}, \bibinfo {author}
  {\bibfnamefont {C.}~\bibnamefont {Heimann}}, \bibinfo {author} {\bibfnamefont
  {H.}~\bibnamefont {Kalinowsky}}, \ and\ \bibinfo {author} {\bibfnamefont
  {W.}~\bibnamefont {Jhe}},\ }\bibfield  {title} {\enquote {\bibinfo {title}
  {Precision mass spectroscopy of the antiproton and proton using
  simultaneously trapped particles},}\ }\href@noop {} {\bibfield  {journal}
  {\bibinfo  {journal} {Physical Review Letters}\ }\textbf {\bibinfo {volume}
  {82}},\ \bibinfo {pages} {3198} (\bibinfo {year} {1999})}\BibitemShut
  {NoStop}%
\bibitem [{\citenamefont {Wagner}\ \emph {et~al.}(2012)\citenamefont {Wagner},
  \citenamefont {Schlamminger}, \citenamefont {Gundlach},\ and\ \citenamefont
  {Adelberger}}]{wagner2012torsion}%
  \BibitemOpen
  \bibfield  {author} {\bibinfo {author} {\bibfnamefont {T.~A.}\ \bibnamefont
  {Wagner}}, \bibinfo {author} {\bibfnamefont {S.}~\bibnamefont
  {Schlamminger}}, \bibinfo {author} {\bibfnamefont {J.}~\bibnamefont
  {Gundlach}}, \ and\ \bibinfo {author} {\bibfnamefont {E.~G.}\ \bibnamefont
  {Adelberger}},\ }\bibfield  {title} {\enquote {\bibinfo {title}
  {Torsion-balance tests of the weak equivalence principle},}\ }\href@noop {}
  {\bibfield  {journal} {\bibinfo  {journal} {Classical and Quantum Gravity}\
  }\textbf {\bibinfo {volume} {29}},\ \bibinfo {pages} {184002} (\bibinfo
  {year} {2012})}\BibitemShut {NoStop}%
\bibitem [{\citenamefont {Barnes}(2012)}]{barnes2012gravitational}%
  \BibitemOpen
  \bibfield  {author} {\bibinfo {author} {\bibfnamefont {J.~E.}\ \bibnamefont
  {Barnes}},\ }\bibfield  {title} {\enquote {\bibinfo {title} {Gravitational
  softening as a smoothing operation},}\ }\href@noop {} {\bibfield  {journal}
  {\bibinfo  {journal} {Monthly Notices of the Royal Astronomical Society}\
  }\textbf {\bibinfo {volume} {425}},\ \bibinfo {pages} {1104--1120} (\bibinfo
  {year} {2012})}\BibitemShut {NoStop}%
\end{thebibliography}%

\newpage 

\appendix

\section{Detailed anomalous heating calculation}
\label{appendix-heating}

In this appendix, we provide a detailed calculation of the anomalous heating effects present in our strongly incoherent model. In particular, we consider the change in the momentum variance
\be
\label{dtvar}
\braket{\partial_t (\Delta \mb{p}_i)^2} = \braket{\partial_t (\mb{p}_i^2)} - \braket{\partial_t \mb{p}_i} \cdot \braket{\mb{p}_i}.
\ee
We will continue to use the notation in the main text in which we use a hat to distinguish position operators $\hat{\mb{x}}$ from POVM outcomes $\mb{x}$, while other operators like $\mb{p}$ have no hats. For simplicity we will assume that the POVM operators $P(\mb{x})$ are Hermitian, as for example in our explicit Gaussian example \eqref{gaussianpovm}. 

From the usual adjoint Lindblad equation, the first term is
\begin{align}
\label{dpdt3}
\braket{\partial_t (\mb{p}_i^2)} = i \braket{[H,\mb{p}_i^2]}  - \frac{1}{2} \sum_j \gamma_j \int d^3\mb{x} \braket{ \{P_j^2(\mb{x}),\mb{p}_i^2\}} - 2 \braket{E_j^\dag(\mb{x}) \mb{p}_i^2 E_j(\mb{x})}.
\end{align}
Consider first the terms $i=j$. We have
\be
\braket{E_i^\dag(\mb{x}) \mb{p}_i^2 E_i(\mb{x})} = \braket{P_i(\mb{x}) \mb{p}_i^2 P_i(\mb{x})}
\ee
since all of the $U_{k \neq i}$ terms in the jump operator $E_i(\mb{x}) = P_i(\mb{x}) \prod_{k \neq i} U_k(\mb{x}-\hat{\mb{x}}_k)$ commute through. Notice that for a function $f = f(\hat{\mb{x}})$ of the position operator only, we have $[f(\hat{\mb{x}}), [f(\hat{\mb{x}}), \mb{p}^2]] = -2 [\nabla f(\hat{\mb{x}})]^2$. Thus the two terms in the integral \eqref{dpdt3} combine to a term of the form
\begin{align}
\left\{ P^2, \mb{p}^2 \right\} - 2 P \mb{p}^2 P = P [P,\mb{p}^2] + [\mb{p}^2,P] P = -2 \left( \nabla P(\mb{x}) \right)^2,
\end{align}
and so the $i=j$ term in \eqref{dpdt3} is
\be
\gamma_i  \int d^3\mb{x} \braket{\left(\nabla P_i(\mb{x}) \right)^2}.
\ee
For the $j \neq i$ terms, the final term in the integral can be computed using commutators. We have
\begin{align}
\begin{split}
E_j^\dag(\mb{x}) \mb{p}_i^2 E_j(\mb{x}) & = P_j(\mb{x}) \prod_{k \neq j} U_k^\dag(\mb{x}-\hat{\mb{x}}_k) \mb{p}_i^2 \prod_{\ell \neq j} U_{\ell}(\mb{x}-\hat{\mb{x}}_{\ell}) P_j(\mb{x}) \\
& = P_j^2(\mb{x}) U_i^\dag(\mb{x}-\hat{\mb{x}}_i) \mb{p}_i^2 U_i(\mb{x}-\hat{\mb{x}}_i) \\
& = P_j^2(\mb{x}) \Big[ \mb{p}_i^2 + \eta_{ij} \left\{ \mb{p}_i, \nabla \phi(\mb{x}-\hat{\mb{x}}_i) \right\} + \eta_{ij}^2 \left( \nabla \phi(\mb{x}-\hat{\mb{x}}_i) \right)^2 \Big]
\end{split}
\end{align}
The first term cancels out with the first term under the integral in \eqref{dpdt3}. Thus in total, we obtain
\begin{align}
\begin{split}
\label{final}
& \braket{\partial_t(\mb{p}_i)^2} = i \braket{[H,\mb{p}_i^2]} + v^2 m_i  \int d^3\mb{x} \braket{\left(\nabla P_i(\mb{x}) \right)^2} \\
& -\frac{1}{2} \sum_{j \neq i} \int d^3\mb{x} \Big[ G_N m_i m_j \Braket{P_j^2(\mb{x}) \left\{ \nabla \phi(\mb{x}-\hat{\mb{x}}_i), \mb{p}_i \right\}} + \frac{G_N^2 m_i^2 m_j}{v^2} \Braket{P_j^2(\mb{x})\left( \nabla \phi(\mb{x}-\hat{\mb{x}}_i) \right)^2} \Big].
\end{split}
\end{align}
Here we used Eq. \eqref{vsigma} to write the explicit coupling constants in the gravitational case. Equation \eqref{final} is an exact result. We now want to estimate its effects in the limits relevant to the experiments discussed in the main text. 

The first non-Hamiltonian term is easy to understand: it is a pure back-action noise coming from the position measurements on the $i$th mass, as in Eq. \eqref{dpdtba}. This can be estimated as $\partial_x P \sim P/\sigma$, and so this terms becomes $\sim \gamma_i \int d^3x \braket{P_i^2(x)}/\sigma^2 = \gamma_i/\sigma^2$ using completeness of the POVM. 

The term of $\mathcal{O}(G_N)$ represents the average force. In particular, consider the limit in which the $j \neq i$ particles are well-separated from the $i$th particle, and the joint state is unentangled. Furthermore, assume the state of $i$ is localized enough that the force $\nabla \phi \approx \mb{F}_0$ acting on it is approximately spatially constant. Then the expectation value factorizes as $\braket{P_j^2(\mb{x})} \mb{F}_0 \cdot \braket{\mb{p}_i}$, which exactly cancels the term $\braket{\partial_t \mb{p}_i} \cdot \braket{\mb{p}_i}$ in \eqref{dtvar}.

Finally, the last term represents shot noise from the gravitational interaction itself. Again we assume that the particles are well-separated and un-entangled. The expectation value factors then as $\braket{P_j^(\mb{x})} \braket{ (\nabla \phi)^2}$. Together with the result of the previous paragraph, and assuming the Hamiltonian itself preserves the variance, this means that \eqref{final} reduces to
\begin{align}
\braket{\partial_t(\Delta \mb{p}_i)^2} = v^2 m_i  \int d^3\mb{x} \braket{\left(\nabla P_i(\mb{x}) \right)^2}  - \sum_{j \neq i} \frac{G_N^2 m_i^2 m_j}{2 v^2} \int d^3\mb{x} \Braket{P_j^2(\mb{x})} \Braket{ \left( \nabla \phi(\mb{x}-\hat{\mb{x}}_i) \right)^2}.
\end{align}
This is the expression used to estimate various bounds in Sec. \ref{sec-gravity}.

\section{Visibility calculations}
\label{appendix-visibility}

Here we give detailed computations of the atomic coherence $\braket{\sigma_-}$ and visibility $V = |\braket{\sigma_-}|$ signatures, in both the dipole and quadrupole limits, discussed in the main text. In the general geometry of Fig. \ref{fig-geometries}, we will assume that the separation $\mb{d}$ is large compared to the displacements $\mb{x}_m, \mb{x}_M$ from equilibrium of the two massive objects. The total distance between the objects $\mb{X} = \mb{d} - (\mb{x}_m - \mb{x}_M)$ then admits a multipole expansion of the potential:
\be
\label{multipole}
\phi(\mb{X}) = \frac{1}{d} \left( 1 - \frac{d_i x^i}{d^3} + \frac{(d_i d_j - 3 \delta_{ij} d^2) x^i x^j}{d^5} + \cdots \right),
\ee
where $\mb{x} := \mb{x}_m - \mb{x}_M$. The dipole term $\sim d_i x^i = d_i (x_m^i + x_M^i)$ is a sum of the two object positions, and therefore cannot lead to any entanglement, regardless of the gravitational model. This term (or specifically the $d_i x_m^i$ part) is responsible for the classical phase shift in the types of interferometry experiments discussed around \eqref{classical-main}.

Now we turn to some effects of the quadrupole terms. In standard quantized gravity, these lead to entanglement; in our strongly incoherent gravity model, they do not. More importantly for us, they lead to very different coherence signatures. We make the approximation that the position resolution $\sigma$ is sufficiently imprecise such that the atoms are not kicked out of the trap when they are measured, but also precise enough that the two trap locations can be detected. In this limit the position measurement on the atom can be approximated by two outcomes
\be
P_L = \frac{\ket{L} \bra{L}}{\sqrt{2}}, P_R = \frac{\ket{R} \bra{R}}{\sqrt{2}},
\ee
where the normalization is such that $\sum_{i \in \{ L,R \}} P_i^\dag P_i = 1$. The feedback force operator on the oscillator will not be important in what follows, since we assume it is very heavy $M \gg m$. The position operator on the oscillator produces a continuous outcome $\mb{x}_M$. The quadrupole part of the feedback force on the atom given outcome $\mb{x}_M$ is
\be
U = \exp \left( -i \theta z_M \sigma_z \right), \ \ \ \theta = \frac{2 G_N m \ell}{v^2 d^3},
\ee
wheree have taken the approximation that we can focus motion on the $y,z$ axes only, following the discussion in Sec. \ref{sec-tests}.

Using these expressions in the adjoint version of \eqref{lindblad}, we find the equation of motion for $\braket{\sigma_-(t)}$:
\be
\label{sigmadot}
\braket{\dot{\sigma}_-} = - \gamma_a \braket{\sigma_-} - 2 \gamma_0 \left[ \braket{\sigma_-} - \braket{\mathcal{P} \sigma_-} \right],
\ee
where as above $\gamma_m = v^2 m$ is the rate of pure atomic dephasing from its position measurements, $\gamma_M = v^2 M$, and $\mathcal{P}$ is an operator on the oscillator given by
\be
\label{calP}
\mathcal{P} = \int d^3\mb{x} P^\dag(\mb{x}) P(\mb{x}) e^{-i \theta z_M \sigma_z} = e^{- i \theta \hat{z}_M} e^{- \theta^2 \sigma^2}.
\ee
The second equality follows by trivially integrating over the $x,y$ axes, then shifting the remaining integral $z_m \to z_M - \hat{z}_M$ (assuming the Gaussian POVM) and using $\sigma_z^2 = 1$. We are mainly interested in the time derivative of the visibility $V = | \braket{\sigma_-}|$. Using \eqref{sigmadot}, this is
\be
\dot{V} = - \gamma_m - 2 \gamma_M \left[ 1 - {\rm Re} \left( \frac{\braket{\mathcal{P} \sigma_-}}{\braket{\sigma_-}} \right) \right].
\ee
We see that the visibility is decreasing unless the final term is sufficiently large. Specifically, revival of the visibility will occur when 
\be
\label{revival}
{\rm Re} \left( \frac{\braket{\mathcal{P} \sigma_-}}{\braket{\sigma_-}} \right) \gtrsim 1
\ee
in the relevant parameter regime $\gamma_m/\gamma_M = m/M \ll 1$.

Our next task is to estimate when the revival condition \eqref{revival} can be satisfied in our experiment. If this were possible, observing the revival would not directly rule out our strongly incoherent gravity model. However, as we now demonstrate, the visibility is always decreasing given the initial states relevant to the proposal \cite{Carney:2021yfw}.

Consider first an initial product state $\rho = \rho_m \otimes \rho_M$. Then we have $\braket{\mathcal{P} \sigma_-} = \braket{\mathcal{P}} \braket{\sigma_-}$, and so the revival will occur if $\braket{\mathcal{P}} \gtrsim 1$. But $|\mathcal{P}| \leq 1$ immediately from its definition \eqref{calP}. Moreover, since the strongly incoherent gravity channel is non-entangling, an initially unentangled state will remain unentangled. Thus any initial product state will have strictly decreasing visibility, and no revival will ever occur. In particular, observation of atomic revival in the ``un-boosted'' protocol of \cite{Carney:2021yfw} would rule out the entire strongly incoherent gravity model.

Next, consider starting with an initial entangled state which ``boosts'' the collapse and revival dynamics by maximizing the quantum Fisher information for observing the incoherent processes. It is possible to find such states where the revival can occur, but the initially pre-entangled states suggested in \cite{Carney:2021yfw} do not lead to revival. In particular, in the ``boosted'' protocol where we start with an initially entangled state generated by a non-gravitational coupling, we have the initial state\footnote{This can be derived by taking $\rho_0 = (\ket{L} + \ket{R}) (\bra{L} + \bra{R}) \otimes \rho_T/2$ and evolving under the interaction Hamiltonian, see Eq. (7) in \cite{Carney:2021yfw}.}
\begin{align}
\rho = \frac{1}{2} \sum_{ij \in \{ L,R \}} \ket{i} \bra{j} \otimes D_{i} \rho_{T} D_j^{\dag},
\end{align}
where $D_i = D(\beta_i)$ is a displacement operator with amplitude $\beta_i = \pm 2 \lambda' = 2 g'/\omega$ in the notation of \cite{Carney:2021yfw}, with $g'$ the non-gravitational coupling generating the initial entanglement, and $\rho_{T}$ is the thermal density matrix for the oscillator. In this state, we have\footnote{Here we used the braiding relation $D(\alpha) D(\beta) = e^{(\alpha \beta^* - \alpha^* \beta)/2}$ and inner product $\braket{\beta | \alpha} = e^{-|\alpha - \beta|^2/2} e^{(\alpha \beta^* - \alpha^* \beta)/2}$.}
\begin{align}
\begin{split}
\braket{\sigma_-} & = \frac{1}{2} \tr D_R \rho_T D_L^\dag \\
& = \frac{1}{2} \int d^2\alpha P_T(\alpha) \braket{\alpha | D_L^\dag D_R | \alpha} \\
& = \frac{1}{2} \int d^2\alpha P_T(\alpha) \braket{\alpha | D^\dag(\beta) D(-\beta) | \alpha}.
\end{split}
\end{align}
Here $P_T(\alpha) = e^{-|\alpha|^2/\bar{n}}/\pi \bar{n}$ is the thermal distribution and we used the fact that $\beta_R = - \beta_L \equiv -\beta$. The inner product is
\begin{align}
\begin{split}
\braket{\alpha | D^\dag(\beta) D(-\beta) | \alpha} &  = e^{-\alpha^* \beta + \alpha \beta^*} \braket{ \alpha + \beta | \alpha - \beta} \\
& = e^{-2|\beta|^2} e^{-2 \alpha^* \beta + 2 \alpha \beta^*}.
\end{split}
\end{align}
We then need the thermal average
\be
\int d^2\alpha P_T(\alpha) e^{-2 \alpha^* \beta + 2 \alpha \beta^*} = e^{-4 \bar{n} |\beta|^2},
\ee
so in total we obtain
\be
\braket{\sigma_-} = \frac{1}{2} e^{-(4\bar{n}+2) | \beta|^2}.
\ee
Meanwhile, the numerator of Eq.~\ref{revival} is
\begin{align}
\begin{split}
\braket{\mathcal{P} \sigma_-} & = \frac{1}{2} \tr \mathcal{P} D_R \rho_T D_L^\dag \\
& = \frac{1}{2} \int d^2\alpha P_T(\alpha) \braket{\alpha | D^\dag(\beta) \sqrt{\mathcal{P}} \sqrt{\mathcal{P}} D(-\beta) | \alpha}.
\end{split}
\end{align}
This is more painful to evaluate because of the $\mathcal{P}$ operator. We used the square root to make this more symmetric, where $\sqrt{\mathcal{P}} = e^{-\theta^2 \sigma^2/2} e^{-i \theta \hat{z}_M/2}$ [see Eq. \eqref{calP}]. One strategy to evaluate this is to note that $e^{-i \theta \hat{z}_M}$ generates a momentum translation $p \to p + \theta$, so it acts on coherent states as $e^{-i \theta \hat{z}_M} \ket{\alpha} = \ket{\alpha + i \theta z_0}$, where $z_0 = \sqrt{1/2 M \omega}$ is the zero-point amplitude of the oscillator. This can be seen explicitly by comparing to the displacement operator $D$ with an imaginary argument. The inner product can therefore be reduced to
\begin{align}
\begin{split}
\braket{\alpha | D^\dag(\beta) \sqrt{\mathcal{P}} \sqrt{\mathcal{P}} D(-\beta) | \alpha} & =  e^{-\alpha^* \beta + \alpha \beta^*} \braket{ \alpha + \beta | \sqrt{\mathcal{P}} \sqrt{\mathcal{P}} | \alpha - \beta} \\
& = e^{-\theta^2 \sigma^2} e^{-\alpha^* \beta + \alpha \beta^*} \braket{ \alpha + \beta + i \theta z_0/2 | \alpha - \beta - i \theta z_0/2} \\
& = e^{-\theta^2 \sigma^2-\theta^2 z_0^2/2} e^{-2 |\beta|^2 + 2 \theta {\rm Im} \beta} e^{-2\alpha^* \beta + 2\alpha \beta^* -i \theta z_0 {\rm Re} \alpha}.
\end{split}
\end{align}
The first two terms here are independent of the thermal averaging. The third term gives the thermal value
\begin{align}
\int d^2\alpha P_T(\alpha) e^{-2\alpha^* \beta + 2\alpha \beta^* -i \theta z_0 {\rm Re} \alpha} = e^{-4 \bar{n} ( |\beta|^2 + \theta z_0 {\rm Im} \beta + \theta^2 z_0^2/4)}.
\end{align}
Putting this together, we get
\begin{align}
\begin{split}
\braket{\mathcal{P} \sigma_-} & = \frac{1}{2} e^{-\theta^2 \sigma^2-\theta^2 z_0^2/2} e^{-2 |\beta|^2 + 2 \theta {\rm Im} \beta} e^{-4 \bar{n} ( |\beta|^2 + \theta z_0 {\rm Im} \beta + \theta^2 z_0^2/4)} \\
& = \frac{1}{2} e^{-(4\bar{n}+2) |\beta|^2} e^{-\theta^2 \sigma^2 - (\bar{n}+1/2) \theta^2 z_0^2 - (4\bar{n}-2) \theta z_0 {\rm Im} \beta}.
\end{split}
\end{align}
Finally, we can now divide by $\braket{\sigma_-}$, to obtain
\be
\frac{\braket{\mathcal{P} \sigma_-}}{\braket{\sigma_-}} = e^{-\theta^2 \sigma^2 - (\bar{n}+1/2) \theta^2 z_0^2 - (4\bar{n}-2) \theta z_0 {\rm Im} \beta}.
\ee
The exponent is manifestly negative except for the final term proportional to ${\rm Im} \beta = \Delta p$, where $\Delta p = g'/\omega' z_0$ is the initial relative momenta of the two conditional oscillator states in the boosted protocol. Revival can occur only when this term is at least as large as the others. But in particular, this requires that $\bar{n} \lesssim 1/2$, i.e., the oscillator has to be initialized (before the boost) very near the ground state. In any non-trivially thermal state, revival will not occur, at least initially. This is in stark contrast to the prediction of quantized gravity, which was shown in \cite{Carney:2021yfw} to lead to visibility revival with the same initial conditions. 

Finally, we should be clear how this model is able to reproduce standard interferometry results but differs from the kind of classical model studied in \cite{Hosten:2021biw}. Consider a ``large'' source mass $M$ placed some distance $d$ from an atom of mass $m$ superposed in a state like $\ket{L} + \ket{R}$. As above, we approximate the interaction potential as
\be
V_{\rm int} \approx \frac{G_N M m x \ell \sigma_z}{d^3}
\ee
where $\sigma_z$ measures the two atom locations. 

The ``classical'' model considered in \cite{Hosten:2021biw} is to take the source mass position $x = x_{\rm cl}$ as a classical c-number, possibly averaged over thermal states. In the simplest case, we can imagine that $M$ is just stationary. The atom state evolves as
\be
\ket{L} + \ket{R} \to e^{-i \phi(t)} \ket{L} + e^{+i \phi(t)} \ket{R}, \ \ \phi(t) = G_N M m \ell t/L^3,
\ee
so we have
\be
\label{classical}
\braket{\sigma_-(t)} = e^{-2 i \phi(t)} \braket{\sigma_-(0)}.
\ee
This explains the usual COW-type experiment (Colella-Overhauser-Werner), where a superposed atom (or neutron) is placed above the fixed Earth, and one can see interference fringes with contrast proportional to $G_N$. On the other hand, the visibility $V(t) = | \braket{\sigma_-(t)}| \equiv V(0)$ is constant in time. This calculation can be extended to allow $M$ to be harmonic and in a thermal state, and one finds revival of the visibility. 

In contrast, our model requires that we treat the oscillator as a quantum system. To compare to the above classical model, we can imagine that $M$ is in a very well-localized state, for example, the ground state in a very tight trap $\psi_M(x) = N_M e^{-(x-x_{\rm cl})^2/2 \sigma_M^2}$ with $\sigma_M \to 0$. Eq. \eqref{sigmadot} then gives (ignoring the atomic dephasing $\gamma_a \to 0$)
\begin{align}
\begin{split}
\braket{\dot{\sigma}_-} & = - 2 \gamma_o \left[ 1 - \int dx \braket{P^2(x) e^{-i \theta \hat{x}}} \right] \braket{\sigma_-} \\
& = - 2 \gamma_o \left[ 1 - e^{-\theta^2 \sigma_M^2/2} e^{-i \theta x_{\rm cl}} \right] \braket{\sigma_-} \\
& \approx -2 \gamma_o \left[ i \theta x_{\rm cl} + \frac{\theta^2}{2} (\sigma_M^2 + x_{\rm cl}^2) \right] \braket{\sigma_-} \\
& = -2 \frac{G_N M m \ell}{L^3} \left[ i x_{\rm cl} + \frac{G_N M m \ell}{2 v^2 L^3} (\sigma_M^2 + x_{\rm cl}^2) \right] \braket{\sigma_-}.
\end{split}
\end{align}
We used $\theta = G_N m \ell/v^2 L^3$ to write the last line explicitly. We see that the term in brackets produces two key effects. At leading order in the small coupling, we get an oscillating term---precisely the same as Eq. \eqref{classical}. Note that the powers of $v^2$ cancelled, as they must. At second order, however, there is a pure loss of the contrast. In particular, the visibility goes as
\be
\dot{V} = - \left( \frac{G_N M m \ell}{ L^3}\right)^2 \frac{1}{v^2}.
\ee
This could again be extended to a harmonic, thermal oscillator, but the basic difference with the classical model of \cite{Hosten:2021biw} should be clear from the simple stationary example. This loss of visibility is due to the gravitational shot noise.

\end{document}